\documentclass[12pt,oneside,a4paper]{article}

\usepackage{color}
\usepackage{epsfig}
\usepackage{amsmath}
\usepackage{amssymb}
\usepackage{bm}

\begin{document}

\begin{center}

\vspace{1cm}

{\bf \large On Functional and Holographic Renormalization Group Methods in Stochastic Theory of Turbulence}

\vspace{1cm}

{\bf \large S.L. Ogarkov}

\vspace{0.5cm}

{\it The Center of Fundamental and Applied Research,}

{\it Dukhov Research Institute of Automatics, Moscow, Russia}

\vspace{1cm}

\abstract{A nonlocal quantum-field model is constructed for the system of hydro\-dynamic equations for
incompressible viscous fluid (the stochastic Navier--Stokes (NS) equation and the continuity equation). This
model is studied by the following two mutually parallel methods: the Wilson--Polchinski functional
renormalization group method (FRG), which is based on the exact functional equation for the generating
functional of amputated connected Green's functions (ACGF), and the Heemskerk--Polchinski holographic
renormaliza\-tion group method (HRG), which is based on the functional Hamilton--Jacobi (HJ) equation for the
holographic boundary action. Both functional equations are equivalent to infinite hierarchies of
integro-differential equations (coupled in the FRG case) for the corresponding families of Green's functions
(GF). The RG-flow equa\-tions can be derived explicitly for two-particle functions. Because the HRG-flow
equation is closed (contains only a two-particle GF), the explicit analytic solutions are obtained for the
two-particle GF (in terms of the modified Bessel functions $I$ and $K$) in the framework of the minimal
holographic model and its simple generalization, and these solutions have a remarkable property of minimal
dependence on the details of the random force correlator (the function of the energy pumping into the system).
The restrictions due to the time-gauged Galilean symmetry present in this theory, the problem of choosing the
pumping function, and some generalizations of the standard RG-flow procedures are discussed in detail. Finally,
the question of whether the HRG-solutions can be used to solve the FRG-flow equation for the two-particle GF
(in particular, the relationship between the regulators in the two methods) is studied.}

\end{center}


\newpage

\tableofcontents{}

\vspace{0.5cm}

\section{Introduction}\label{p1}

The method of functional (nonperturbative, exact) renormalization group (FRG), just as the Schwinger--Dyson
formalism (functional equation), is based on an infinite hierarchy of integro-differential equations for
certain types of Green's functions or vertex functions. The most important principal difficulty encountered in
all of these hierarchies is the fact that they contain the so-called $n,n+1$ problem. This problem is that the
equation for the $n$-particle Green's function contains the $n+1$-particle function in a special kinematics
(for example, with respect to the momenta). And this is the main difficulty encountered when the stochastic and
quantum field theories are constructed on the functional level. A survey of the contemporary state of the art
in this field and a detailed introduction to the FRG method can be found in the wonderful monograph \cite{KBS}
which, on the one side, is written as a pedagogical introduction and, on the other side, describes a great many
of the last advances in the theory and applications of the method. The well-written Chapter~12 in the monograph
\cite{Wipf} and the reviews \cite{BTW}--\cite{YIKIHS} also deserve special attention.

The simplest idea for solving this problem is to try to change the $n+1$th function in a given order $n$
(usually, $n=2$). This scenario is well known in all areas of contemporary physics such as quantum field
theory, theory of stochastic partial differential equations (SPDE), turbulence, solid-state physics, and many
others. The current situation evolves towards a different favorable scenario; namely, it is assumed that the
$n$th (for example, two-particle) function is prescribed and the other functions can be constructed from the
hierarchy by using the ascending recursion in the upward direction. And the most important question then
arises: What functions should be taken as the initial ones?

And the following recently formulated construction can come to the help here. We mean the so-called holographic
renormalization group (HRG) in the form which was developed in the remarkable work \cite{IHJP} (we also note
\cite{TFHLMR} and \cite{VBMGAL}). We note here that the HRG idea itself appeared almost immediately after the
discovery of the AdS/CFT correspondence (the correspondence between the classical field theory in the anti-de
Sitter curved space and the quantum conformal field theory) \cite{Maldacena}--\cite{Natsuume}, but its earlier
versions were more oriented to the non-Abelian theories and the theories of gravity \cite{MFSMTS}. We also note
the attempts to draw parallels between HRG and the method of stochastic quantization in \cite{JHODPJ}.

The main idea of HRG is that a certain action is formulated in a curved space with additional dimensions (the
anti-de Sitter space is most often used for these purposes) and the functional Hamilton--Jacobi (HJ) equation
is then written for this action. The most important advantage of this equation is that the corresponding
hierarchy of integro-differential equations for Green's functions in the HJ theory does not contain the $n,n+1$
problem! So, for example, a generalization of a special Riccati equation is obtained for the two-particle
Green's function, and the solution of this equation contains a wonderful combination of the modified Bessel
functions $I$ and $K$ (an example of the scalar field theory can be found in the recent paper \cite{LMPV}).

The critics often say that if the AdS/CFT correspondence is indeed true, then this means that the
correspondence is based on some "superstring structure"$\!$ connecting the two theories in a complicated way.
And, of course, this imposes strict restrictions on the classes of quantum-field models for which the
correspondence is true. In the set of such models, precisely the (supersymmetric) non-Abelian theories and the
theories of gravity can be distinguished.

At the same time, the HRG predictions for any prescribed theory can always be verified using the FRG. For this,
it is necessary to substitute the HRG hierarchies directly into the FRG hierarchy. In this case, there is a
certain "degree of freedom"$\!$ which means that the relationship between the regulators in the two theories is
not known exactly (this is discussed, for example, in \cite{IHJP}, where some comments on the possible solution
of this problem are given).

But this ambiguity is not a serious drawback of the HRG method. Moreover, this drawback can be treated as an
element of the statement of the problem, and this can even be an advantage. Furthermore, in contrast to the
(quantum) scalar field theory, there are several new "players"$\!$ in the stochastic theory of turbulence
"living"$\!$ in the curved ambient space, and this fact permits obtaining very interesting answers for the
Green's functions.

In particular, one can construct a functional flow weakly depending on the details of the function describing
the energy pumping into the system (the random force correlator in the momentum-frequency representation) and
also exhibiting some other interesting properties. Conversely, one can use the independence property and some
other properties to fix the \emph{conditions of choice}, i.e., the rules for constructing the flow which permit
eliminating any ambiguity in the functional HJ equation. This is a peculiar symmetry which permits fixing
certain quantities in the theory, and such conditions are of principal importance in this paper.

Another important point is the following one: the AdS/CFT correspondence standardly holds in a certain
two-sided limit. And this fact must be taken into account when functions from HRG are substituted into FRG. But
one can play a different game, i.e., verify whether the FRG hierarchy holds if, for example, it is assumed that
the solution for the two-particle function in HRG is exact and what form the remainder has if the answer to
this question is negative.

We further note that the HRG method is much wider than the classical AdS/CFT correspondence. Indeed, the
functional HJ equation can be solved in any metric. It is only required that the metric tensor must have
several properties which are exhibited in the AdS. Such a metric "launches"$\!$ a functional flow, but its
details are already of no importance. In a sense, one can treat the metric as another "pumping function"$\!$,
and utter words "gravity/pumping duality"$\!$ about the duality between the gravity and the energy pumping into
the system.

All this is based on the following simple fact already known in the theory of nonlinear equations for usual
functions of one or several variables: the equations with more complicated nonlinearity in a certain domain of
parameters have solutions "similar"$\!$ to the solutions of equations which are simpler in the sense of
nonlinearity but contain fractional derivatives or more complicated coefficients. If we now add the word
"functional"$\!$, then everything takes its own place.

Summarizing the above, we can draw another positive conclusion. Since the FRG method ($n,n+1$ problem) is very
complicated, the HRG approach can be regarded as a separate, independent, simplified ("toy"$\!$) model. Its
results are of their own importance, because they illustrate both the general properties of the functional flow
and the explicit analytic solutions for various families of Green's functions and vertex functions. The results
obtained in the HRG framework require further analysis through the FRG "prism"$\!$.

Moreover, this does not mean that the "HRG-scenario"$\!$ is complete, because a great number of questions
"remains overboard"$\!$; namely, introduction of composite operators \cite{YIKIHS}, analysis of the correlation
and structure functions corresponding to these operators, consideration of $n$-particle irreducible vertex
functions \cite{CFPP}, and many other problems whose solutions, if obtained by the HRG method, must then be
verified at the level of the corresponding FRG hierarchy.

The goal in this paper is to explain the general FRG and HRG ideas and to demonstrate the practical
possibilities of these methods with an example of a nonlocal quantum-field model of the stochastic theory of
turbulence. The following three sections in this paper deal with the three stages required to attain the goal.
At the beginning of Section~2, we successively derive a nonlocal quantum-field model. Further, in this section,
we present the FRG formalism and introduce the Wilson--Polchinski abstract FRG-flow equation which is satisfied
by the ACGF generating functional. The latter is used in the momentum-frequency representation to derive the
FRG-flow equation for the two-particle Green's function. At the end of this section, we study the Ward
functional identity for the time-gauged Galilean symmetry and its consequences, which play an important role in
the quantum-field model of turbulence considered in the FRG method. At the beginning of Section~3, we present
an introduction to the HRG formalism and formulate the functional Hamilton--Jacobi equation. Further, in the
momentum-frequency representation, we derive the HRG-flow equation for the two-particle boundary Green's
function (BGF). Then, in the framework of the minimal holographic model and its simple generalization, we
construct the solution of this equation explicitly and analytically. We also discuss the properties of the
obtained solution. In conclusion, we consider the problem of choosing a pumping function and make several
generalizations of the RG-flow procedures used in the paper, both from the HRG and FRG sides. In Section~4, we
study whether it is possible to apply the HRG-solutions to construct solutions of the FRG-flow equation for the
two-particle Green's function. Such a "synthesis"$\!$ is an important result, because it permits constructing
some family of Green's functions from the FRG hierarchy by using the ascending recursion in the upward
direction starting from a single known (for example, two-particle) function. The final discussion of all
obtained results is given in the conclusion.

\section{Hydrodynamic equations\\ from the CFT side}\label{p2}

We consider the system of hydrodynamic equations for incompressible viscous fluid \cite{Vasilev}--\cite{AAV}.
This system consists of the Navier--Stokes equation with a random force and the continuity equation (without
external random parameters):
\begin{eqnarray}
\partial_{t}\boldsymbol{v}-\nu\partial^{2}\boldsymbol{v}+\left(\boldsymbol{v}\partial\right)
\boldsymbol{v}+\frac{\partial P}{\rho}-\boldsymbol{f}=\mathbf{0},\quad \partial\boldsymbol{v}=0.
\label{2frgstt1}
\end{eqnarray}
Here $P$ is the pressure, $\rho$ is the density, and $\nu$ is the viscosity. The random force $\boldsymbol{f}$
is assumed to be Gaussian with zero mean and a correlator of the form
\begin{eqnarray}
\left\langle f_{\alpha}\left(t,\boldsymbol{r}\right)f_{\beta}\left(t',\boldsymbol{r}'\right)
\right\rangle=D_{\alpha\beta}\left(\boldsymbol{r}-\boldsymbol{r}'\right)\delta\left(t-t'\right).
\label{2frgstt2}
\end{eqnarray}
We note that the correlator $D$ in expression (\ref{2frgstt2}) is not assumed to be transverse. But as we shall
see below, only the transverse part of $D$ plays a significant role starting from a certain point. In other
words, this statement of the problem is most general.

Let $\mathrm{Eq}\left[\varPhi,\boldsymbol{f}\right]$ be the left-hand sides of the equations in
(\ref{2frgstt1}). We also introduce the following compact notation for the set of fields in the initial
stochastic problem (\ref{2frgstt1})--(\ref{2frgstt2}) and the sources: $\varPhi=\left(\boldsymbol{v},P\right)$
and $J=\left(\,\boldsymbol{j},j_{P}\right)$. For $\mathrm{Eq}\left[\varPhi,\boldsymbol{f}\right]$, we have the
relations
\begin{eqnarray}
\mathrm{Eq}\left[\varPhi,\boldsymbol{f}\right]=\left(\mathrm{eq}_{\alpha}\left[\varPhi\right]-
f_{\alpha},\partial\boldsymbol{v}\right),\quad
\frac{\delta\mathrm{Eq}\left[\varPhi,\boldsymbol{f}\right]}{\delta\varPhi}\equiv
\mathrm{Eq}'\left[\boldsymbol{v}\right].\label{2frgstt3}
\end{eqnarray}

Further, we consider the (nonnormalized) generating functional for unconnected Green's functions of the
quantum-field model for stochastic problem (\ref{2frgstt1})--(\ref{2frgstt2}). In contrast to the standard
procedure of its derivation, which is widely used in the literature (for the subsequent adaptation of the
perturbation theory), \textbf{we do not introduce auxiliary fields} in this paper. In this case, the expression
for the functional $\mathcal{Z}$ (after calculating the functional integral over the random force
$\boldsymbol{f}$ but before calculating the functional integral over the pressure field $P$) becomes
\begin{eqnarray}
\mathcal{Z}\left[J\right]=\int\mathcal{D}\left[\varPhi\right]\mathrm{Det}
\left\{\mathrm{Eq}'\left[\boldsymbol{v}\right]\right\}\delta^{\left(\infty\right)}
\left[\partial\boldsymbol{v}\right]e^{-S\left[\varPhi\right]+\left(J\mid\varPhi\right)}.
\label{2frgstt4}
\end{eqnarray}
In terms of notation (\ref{2frgstt3}), the field-theoretic action of the system contained in (\ref{2frgstt4})
has the form
\begin{eqnarray}
S\left[\varPhi\right]=\frac{1}{2}\int_{t,\boldsymbol{r}}\mathrm{eq}_{\alpha}D^{-1}_{\alpha\beta}\,
\mathrm{eq}_{\beta}.\label{2frgstt5}
\end{eqnarray}
In expression (\ref{2frgstt4}), we also introduce the compact notation for the scalar product of two fields
\begin{eqnarray}
\left(J\mid\varPhi\right)=\int_{t,\boldsymbol{r}}\left[j\left(t,\boldsymbol{r}\right)
\varphi\left(t,\boldsymbol{r}\right)+j'\left(t,\boldsymbol{r}\right)
\varphi'\left(t,\boldsymbol{r}\right)\right].\label{scalprod}
\end{eqnarray}
Further, we consider the generating functional $\mathcal{Z}$ in more detail.

\subsection{Field-theoretic generating functional}

Calculating the functional integral over the pressure field $P$, we obtain the expression for the functional
$\mathcal{Z}$:
\begin{eqnarray}
\mathcal{Z}\left[\,\boldsymbol{j}\right]=\int\mathcal{D}\left[\boldsymbol{v}\right]
\mathrm{Det}\left\{\mathrm{Eq}'\left[\boldsymbol{v}\right]\right\}\delta^{\left(\infty\right)}
\left[\partial\boldsymbol{v}\right]e^{-S^{\left(\perp\right)}\left[\boldsymbol{v}\right]+
\left(\boldsymbol{j}\mid\boldsymbol{v}\right)}.\label{2frgstt6}
\end{eqnarray}
The field-theoretic action of the system contained in (\ref{2frgstt6}) is pure transverse. The longitudinal
part of the action is cancelled out after calculating the integral over $P$. We explicitly write the action and
all of its "construction blocks":
\begin{eqnarray}
S^{\left(\perp\right)}\left[\boldsymbol{v}\right]=\frac{1}{2}\int_{t,\boldsymbol{r}}
v^{\star}_{\alpha}\left(t,\boldsymbol{r}\right)N_{\alpha\beta}\left(\partial\right)
v^{\star}_{\beta}\left(t,\boldsymbol{r}\right).\label{2frgstt7}
\end{eqnarray}
Because the operation $N$ cannot generally be reduced to the $\delta$ function, it follows from
(\ref{2frgstt7}) that the quantum-field model expressed only in terms of the velocity field $\boldsymbol{v}$
(without any auxiliary field $\boldsymbol{v}'$) is \textbf{nonlocal} for stochastic problem
(\ref{2frgstt1})--(\ref{2frgstt2}). At this important point, the results obtained in this paper differ from the
results of studies of the local quantum-field model carried out in \cite{LCBDNW1}--\cite{LCBDNW2}. The field
$\boldsymbol{v}^{\star}$ has the following linear and quadratic parts with respect to~$\boldsymbol{v}$:
\begin{eqnarray}
v^{\star}_{\alpha}\left(t,\boldsymbol{r}\right)=Lv_{\alpha}\left(t,\boldsymbol{r}\right)+
v_{\beta}\left(t,\boldsymbol{r}\right)\partial_{\beta}v_{\alpha}
\left(t,\boldsymbol{r}\right).\label{2frgstt8}
\end{eqnarray}
The diffusion operator is defined with the \emph{plus} sign at the derivative with respect to time
(correspondingly, its transpose, the antidiffusion operator, has the \emph{minus} sign at $\partial_{t}$):
\begin{eqnarray}
L=\partial_{t}-\nu\partial^{2},\quad L^{T}=-\partial_{t}-\nu\partial^{2}.\label{2frgstt9}
\end{eqnarray}
We also use the decomposition of the inverse correlator $D^{-1}$ into the longitudinal and transverse parts
(namely, the latter appears in expression (\ref{2frgstt7}) for the transverse field-theoretic action of the
system):
\begin{eqnarray}
D^{-1}_{\alpha\beta}\left(\partial\right)=
M\left(\partial\right)\varPi^{\left(\parallel\right)}_{\alpha\beta}\left(\partial\right)+
N\left(\partial\right)\varPi^{\left(\perp\right)}_{\alpha\beta}\left(\partial\right).
\label{2frgstt10}
\end{eqnarray}
Alternatively, the transverse part of $N$ can be written as
\begin{eqnarray}
N_{\alpha\beta}\left(\partial\right)=N\left(\partial\right)
\varPi^{\left(\perp\right)}_{\alpha\beta}\left(\partial\right).\label{2frgstt11}
\end{eqnarray}
Finally, for the projection operators $\varPi^{\left(\parallel\right)}$ and $\varPi^{\left(\perp\right)}$ in
the $\partial$-representation, we have
\begin{eqnarray}
\varPi^{\left(\parallel\right)}_{\alpha\beta}\left(\partial\right)=
\partial_{\alpha}\partial^{-2}\partial_{\beta},\quad
\varPi^{\left(\perp\right)}_{\alpha\beta}\left(\partial\right)=
\delta_{\alpha\beta}-\varPi^{\left(\parallel\right)}_{\alpha\beta}\left(\partial\right).
\label{2frgstt12}
\end{eqnarray}

In the framework of the FRG method, the next step is to supplement the theory with the $\varLambda$-deformation
and, therefore, with the so-called \emph{deformed} field-theoretic generating functional
$\mathcal{Z}_{\varLambda}$. After this, it is necessary to explain the properties of deformation and,
therefore, of the boundary conditions for $\mathcal{Z}_{\varLambda}$.

\subsection{Deformed generating functional}

First, according to the standard regularization of the problem \cite{Vasilev}--\cite{AAV} and \cite{ZJ}, we set
the functional determinant $\mathrm{Det}$ in the right-hand side of (\ref{2frgstt6}) equal to~$1$. This is
correct, because we therefore can compare the results of calculations (in the framework of a nonlocal field
theory) with similar results obtained directly by iterations of stochastic problem
(\ref{2frgstt1})--(\ref{2frgstt2}). The answers will coincide. Further, to use the FRG technique, it is
necessary to represent the integrand in the right-hand side of (\ref{2frgstt6}) \ as an exponential function.
For this, we "smooth"$\!$ the $\delta$ functional on an interval of the order of~$\varepsilon$:
\begin{eqnarray}
\delta^{\left(\infty\right)}\left[\partial\boldsymbol{v}\right]\rightarrow
e^{-S^{\left(\parallel\right)}_{\varepsilon}\left[\boldsymbol{v}\right]}.
\label{2frgstt13}
\end{eqnarray}
In this case, an additional (longitudinal) Gaussian term appears in the field-theoretic action of the system:
\begin{eqnarray}
S^{\left(\parallel\right)}_{\varepsilon}\left[\boldsymbol{v}\right]=\frac{1}{2}
\int_{t,\boldsymbol{r}}v_{\alpha}\left(t,\boldsymbol{r}\right)N_{\varepsilon,\alpha\beta}
\left(\partial\right)v_{\beta}\left(t,\boldsymbol{r}\right).\label{2frgstt14}
\end{eqnarray}
In expression (\ref{2frgstt14}), we introduced the (longitudinal) operation~$N_{\varepsilon}$:
\begin{eqnarray}
N_{\varepsilon,\alpha\beta}\left(\partial\right)=N_{\varepsilon}\left(\partial\right)
\varPi^{\left(\parallel\right)}_{\alpha\beta}\left(\partial\right),\quad
N_{\varepsilon}\left(\partial\right)=D^{-1}_{\varepsilon}\left(\partial\right)
\partial^{2}.\label{2frgstt15}
\end{eqnarray}
This operation has the meaning of a regulator. But this regulator is not an FRG-regulator. The parameter
$\varepsilon$ is fixed and does not determine any FRG-flow. Appar\-ent\-ly, in the general case, one can
imagine a situation where $\varepsilon$ is consistent with the FRG-parameter~$\varLambda$ of the flow. But, in
the statement of the problem considered in this paper, we assume that $\varepsilon$ is a small but fixed free
quantity.

Now we equip the theory with an FRG-regulator \cite{KBS}. For this, we supplement the field-theoretic action of
the system with the additional $\varLambda$-dependent (for simplicity, Gaussian) term
\begin{eqnarray}
S_{\varLambda}\left[\boldsymbol{v}\right]=\frac{1}{2}\int_{t,\boldsymbol{r}}
v_{\alpha}\left(t,\boldsymbol{r}\right)R_{\varLambda,\alpha\beta}
\left(\partial\right)v_{\beta}\left(t,\boldsymbol{r}\right).\label{2frgstt16}
\end{eqnarray}
The generating functional $\mathcal{Z}_{\varLambda}$ (together with all other functionals constructed on the
basis of $\mathcal{Z}_{\varLambda}$) then becomes a function of~$\varLambda$:
\begin{eqnarray}
\mathcal{Z}_{\varLambda}\left[\,\boldsymbol{j}\right]=\int\mathcal{D}\left[\boldsymbol{v}\right]
e^{-S^{\left(\perp\right)}\left[\boldsymbol{v}\right]-S^{\left(\parallel\right)}_{\varepsilon}
\left[\boldsymbol{v}\right]-S_{\varLambda}\left[\boldsymbol{v}\right]+
\left(\,\boldsymbol{j}\mid\boldsymbol{v}\right)}.\label{2frgstt17}
\end{eqnarray}
Thus, we obtained the deformed functional $\mathcal{Z}_{\varLambda}$. Its relation to the initial functional
$\mathcal{Z}$ and the general properties of the $\varLambda$-deformation are analyzed in detail in the
framework of the general FRG formalism~\cite{KBS}. Here we do not discuss this analysis, but only successively
write out the explicitly deformed Gaussian propagator of our theory and the family of Green's functions which
can be used in the FRG method, discuss the additional symmetries of the Navier--Stokes field theory, and then
"take a course"$\!$ to the curved ambient space, for example, the anti-de Sitter space.

\subsection{Deformed Gaussian propagator}

The $\varLambda$-dependent (inverse) Gaussian propagator, which was obtained in the preceding subsection, has
the following form in the $\left(t,\boldsymbol{r}\right)$-representation (the $\partial$-representa\-tion):
\begin{eqnarray}
\left[\mathbf{G}^{-1}_{0,\varLambda}\left(\partial_{t},\partial\right)\right]_{\alpha\beta}=
N_{\varepsilon,\alpha\beta}\left(\partial\right)+N_{\alpha\beta}\left(\partial\right)L^{T}L+
R_{\varLambda,\alpha\beta}\left(\partial\right).\label{2frgstt18}
\end{eqnarray}
It is convenient to pass to the $\left(\omega,\boldsymbol{k}\right)$-representation, where the expression for
the $\varLambda$-dependent (inverse) Gaussian propagator becomes
\begin{eqnarray}
\left[\mathbf{G}^{-1}_{0,\varLambda}\left(\omega,\boldsymbol{k}\right)\right]_{\alpha\beta}=
\frac{\varPi^{\left(\parallel\right)}_{\alpha\beta}\left(\boldsymbol{k}\right)}
{G^{\left(\parallel\right)}_{0,\varepsilon}\left(k\right)}+\frac{\varPi^{\left(\perp\right)}_{\alpha\beta}
\left(\boldsymbol{k}\right)}{G^{\left(\perp\right)}_{0,\varLambda}
\left(\omega,k\right)}.\label{2frgstt19}
\end{eqnarray}
The longitudinal and transverse scalar propagators, which appear in expression (\ref{2frgstt19}), have the form
\begin{eqnarray}
G^{\left(\parallel\right)}_{0,\varepsilon}\left(k\right)=\frac{D_{\varepsilon}\left(k\right)}{k^{2}},
\quad G^{\left(\perp\right)}_{0,\varLambda}\left(k\right)=\frac{D\left(k\right)}
{\omega^{2}+\left(\nu k^{2}\right)^{2}+D\left(k\right)R_{\varLambda}\left(k\right)}.\label{2frgstt20}
\end{eqnarray}
In expression (\ref{2frgstt20}), we only assumed that the FRG-regulator $R_{\varLambda}$ was chosen in
transverse form
\begin{eqnarray}
R_{\varLambda,\alpha\beta}\left(\partial\right)=R_{\varLambda}\left(\partial\right)
\varPi^{\left(\perp\right)}_{\alpha\beta}\left(\partial\right),\quad
D\left(\partial\right)=N^{-1}\left(\partial\right).\label{2frgstt21}
\end{eqnarray}
Precisely this statement of the problem is the simplest and most "natural"$\!$.

\subsection{Amputated connected Green's functions}

In the FRG method, it was shown that the generating functional $\mathcal{G}_{\mathrm{ac},\varLambda}$ of the
so-called \emph{amputated connected Green's functions} (ACGF) satisfies a "convenient"$\!$ boundary condition
in the limit as $\varLambda\rightarrow \infty$, where the Gaussian propagator is switched off \cite{KBS}. We
first introduce the compact notation (from now on, $\boldsymbol{u}$ is an argument of the functional
$\mathcal{G}_{\mathrm{ac},\varLambda}$ and the index $\mathrm{ac}$ is omitted):
\begin{eqnarray}
\mathcal{G}_{\mathrm{ac},\varLambda}\left[\boldsymbol{u}\right]\equiv
\mathcal{G}_{\varLambda}\left[\boldsymbol{u}\right],\quad
\mathcal{G}^{\left(1\right)}_{\varLambda}\left[\boldsymbol{u}\right]=
\frac{\delta}{\delta\boldsymbol{u}}\,\mathcal{G}_{\varLambda}\left[\boldsymbol{u}\right],\quad
\mathcal{G}^{\left(2\right)}_{\varLambda}\left[\boldsymbol{u}\right]=
\frac{\delta}{\delta\boldsymbol{u}}\otimes\frac{\delta}{\delta\boldsymbol{u}}\,
\mathcal{G}_{\varLambda}\left[\boldsymbol{u}\right]\ldots\label{3frgstt13}
\end{eqnarray}
As is known, the functional $\mathcal{G}_{\varLambda}$ has the following representation (the second line in
this formula):
\begin{eqnarray}
e^{\mathcal{G}_{\varLambda}\left[\boldsymbol{u}\right]}=\frac{1}{\mathcal{Z}_{0,\varLambda}
\left[\,\boldsymbol{j}=\mathbf{0}\right]}\int\mathcal{D}\left[\boldsymbol{v}\right]
e^{-S_{0,\varLambda}\left[\boldsymbol{v}\right]-S_{1}
\left[\boldsymbol{v}+\boldsymbol{u}\right]}=\nonumber\\
=e^{\frac{1}{2}\left(\frac{\delta}{\delta\boldsymbol{u}}\mid
\mathbf{G}_{0,\varLambda}\frac{\delta}{\delta\boldsymbol{u}}\right)}
e^{-S_{1}\left[\boldsymbol{u}\right]}.\label{3frgstt14}
\end{eqnarray}
It follows from this representation that in the limit $\varLambda\rightarrow \infty$, where the Gaussian
propagator is switched off, the functional $\mathcal{G}_{\varLambda}$ is equal to the part (with the opposite
sign) of the initial total action of the system which corresponds to the interaction:
\begin{eqnarray}
\mathcal{G}_{\varLambda}\left[\boldsymbol{u}\right]\rightarrow -S_{1}\left[\boldsymbol{u}\right]
\,\,\,\textrm{for}\,\,\, \varLambda\rightarrow \infty.\label{3frgstt15}
\end{eqnarray}
The boundary condition (\ref{3frgstt15}) is perfectly appropriate for the exact analysis and can be used as an
element for constructing some ansatz. We now consider the FRG-flow equation for $\mathcal{G}_{\varLambda}$.

\subsection{Abstract Wilson--Polchinski FRG-flow equation}

The abstract FRG-flow equation for the generating functional $\mathcal{G}_{\varLambda}$ is obtained by
differentiating representation (\ref{3frgstt14}) with respect to~$\varLambda$. As a result, we have
\begin{eqnarray}
\partial_{\varLambda}\mathcal{G}_{\varLambda}\left[\boldsymbol{u}\right]=\frac{1}{2}
\left(\mathcal{G}^{\left(1\right)}_{\varLambda}\left[\boldsymbol{u}\right]\,\,\Big|
\left[\partial_{\varLambda}\mathbf{G}_{0,\varLambda}\right]
\mathcal{G}^{\left(1\right)}_{\varLambda}\left[\boldsymbol{u}\right]\right)
+\frac{1}{2}\mathrm{Tr}\left(\left[\partial_{\varLambda}\mathbf{G}_{0,\varLambda}\right]
\mathcal{G}^{\left(2\right)}_{\varLambda}\left[\boldsymbol{u}\right]\right).\label{3frgstt16}
\end{eqnarray}
Abstract equation (\ref{3frgstt16}) can be rewritten in terms of superindices as
\begin{eqnarray}
\partial_{\varLambda}\mathcal{G}_{\varLambda}\left[\boldsymbol{u}\right]=
\frac{1}{2}\left[\partial_{\varLambda}\mathbf{G}_{0,\varLambda}\right]_{ab}
\left[\mathcal{G}^{\left(1\right)}_{\varLambda,a}\left[\boldsymbol{u}\right]
\mathcal{G}^{\left(1\right)}_{\varLambda,b}\left[\boldsymbol{u}\right]+
\mathcal{G}^{\left(2\right)}_{\varLambda,ab}\left[\boldsymbol{u}\right]\right].\label{3frgstt17}
\end{eqnarray}
Equation (\ref{3frgstt16}) for the generating functional $\mathcal{G}_{\varLambda}$ is identical to the
equation for the $\varLambda$-dependent \emph{Wilsonian effective action}, which was first derived by
Polchinski~\cite{Polchinski}. For this reason, the exact abstract FRG-flow equation (\ref{3frgstt16}) is
sometimes called the \emph{Wilson--Polchinski equation}.

We now represent equation (\ref{3frgstt16}) as an infinite hierarchy of coupled integro-differential equations
for ACGF (here we explicitly write only one equation of the hierarchy). The Green's functions
$\mathcal{G}^{\left(n\right)}_{\varLambda,a_{1}\ldots a_{n}}$ with~$n$ external legs are determined in terms of
the functional derivatives $\mathcal{G}_{\varLambda}$ with respect to the field $\boldsymbol{u}$ on the zero
field configurations:
\begin{eqnarray}
\mathcal{G}^{\left(n\right)}_{\varLambda,a_{1}\ldots a_{n}}\left[\boldsymbol{u}=\mathbf{0}\right]
\equiv\mathcal{G}^{\left(n\right)}_{\varLambda,a_{1}\ldots a_{n}}.\label{3frgstt18}
\end{eqnarray}
It is easy to obtain the exact FRG-flow equation for the two-particle ACGF. This equation reads
\begin{eqnarray}
\partial_{\varLambda}\mathcal{G}^{\left(2\right)}_{\varLambda,ab}=\left[\partial_{\varLambda}
\mathbf{G}_{0,\varLambda}\right]_{cd}\left[\mathcal{G}^{\left(2\right)}_{\varLambda,ac}
\mathcal{G}^{\left(2\right)}_{\varLambda,bd}+\frac{1}{2}
\boxed{\mathcal{G}^{\left(4\right)}_{\varLambda,abcd}}\,\right].\label{3frgstt19}
\end{eqnarray}
Obviously, equation (\ref{3frgstt19}) contains the four-particle amputated function. Continuing the derivation,
we see that the equation for the four-particle Green's function contains the six-particle function, etc. This
$n,n+2$ problem is the main problem of the FRG method, and to solve this problem is the key point when any
physical problem is solved by an appropriate method. Let us consider this problem in the momentum-frequency
representation.

\subsection{Momentum-frequency representation}

Due to the translation invariance of our system, it is convenient to work in the momentum-frequency space. The
momentum-frequency representation is a "natural representation"$\!$, because all expressions have the simplest
and most visual form in this representation. It is also convenient, in the set of all ACGF, explicitly to
separate the Dirac $\delta$ function which expresses the momentum-frequency conservation law:
\begin{eqnarray}
\mathcal{G}^{\left(n\right)}_{\varLambda,a_{1}\ldots a_{n}}
=\left(2\pi\right)^{D+1}\delta\left(\varOmega\right)\delta^{(D)}\left(\boldsymbol{K}\right)
\mathcal{G}^{\left(n\right)}_{\varLambda,\alpha_{1}\ldots\alpha_{n}}
\left(\omega_{1},\boldsymbol{k}_{1},\ldots,\omega_{n},\boldsymbol{k}_{n}\right).\label{3frgstt20}
\end{eqnarray}
Here we introduce the notation for the total frequency and momentum:
\begin{eqnarray}
\varOmega=\omega_{1}+\ldots+\omega_{n},\quad
\boldsymbol{K}=\boldsymbol{k}_{1}+\ldots+\boldsymbol{k}_{n}.\label{3frgstt21}
\end{eqnarray}

The exact FRG-flow equation (\ref{3frgstt19}) for the two-particle Green's function in the momentum-frequency
representation becomes
\begin{eqnarray}
\partial_{\varLambda}\mathcal{G}^{\left(2\right)}_{\varLambda,\alpha\beta}
\left(\omega,\boldsymbol{k}\right)=\left[\partial_{\varLambda}
\mathbf{G}_{0,\varLambda}\left(\omega,\boldsymbol{k}\right)\right]_{\gamma\delta}
\mathcal{G}^{\left(2\right)}_{\varLambda,\alpha\gamma}\left(\omega,\boldsymbol{k}\right)
\mathcal{G}^{\left(2\right)}_{\varLambda,\beta\delta}\left(-\omega,-\boldsymbol{k}\right)
+\nonumber\\
+\frac{1}{2}\int_{\varkappa,\boldsymbol{q}}\left[\partial_{\varLambda}
\mathbf{G}_{0,\varLambda}\left(\varkappa,\boldsymbol{q}\right)\right]_{\gamma\delta}
\mathcal{G}^{\left(4\right)}_{\varLambda,\alpha\beta\gamma\delta}
\left(\omega,\boldsymbol{k},\varkappa,\boldsymbol{q}\right).\label{3frgstt22}
\end{eqnarray}
In equation (\ref{3frgstt22}), we introduced the compact notation for the Green's functions in a special
kinematics:
\begin{eqnarray}
\mathcal{G}^{\left(2\right)}_{\varLambda,\alpha\beta}\left(\omega,\boldsymbol{k}\right)\equiv
\mathcal{G}^{\left(2\right)}_{\varLambda,\alpha\beta}\left(\omega,\boldsymbol{k},
-\omega,-\boldsymbol{k}\right),\nonumber\\
\mathcal{G}^{\left(4\right)}_{\varLambda,\alpha\beta\gamma\delta}\left(\omega,\boldsymbol{k},
\varkappa,\boldsymbol{q}\right)\equiv\mathcal{G}^{\left(4\right)}_{\varLambda,\alpha\beta\gamma\delta}
\left(\omega,\boldsymbol{k},-\omega,-\boldsymbol{k},\varkappa,\boldsymbol{q},
-\varkappa,-\boldsymbol{q}\right).\label{3frgstt23}
\end{eqnarray}

We now project the flow equation (\ref{3frgstt22}) on the transverse direction and obtain
\begin{eqnarray}
\partial_{\varLambda}\mathcal{G}^{\left(\perp\right)}_{\varLambda}\left(\omega,\boldsymbol{k}\right)=
\left[\partial_{\varLambda}G^{\left(\perp\right)}_{0,\varLambda}\left(\omega,k\right)\right]
\left[\mathcal{G}^{\left(\perp\right)}_{\varLambda}\left(\omega,\boldsymbol{k}\right)\right]^{2}
+\nonumber\\
+\frac{1}{2}\int_{\varkappa,\boldsymbol{q}}
\left[\partial_{\varLambda}G^{\left(\perp\right)}_{0,\varLambda}
\left(\varkappa,q\right)\right]\varPi^{\left(\perp\right)}_{\alpha\beta}\left(\boldsymbol{k}\right)
\varPi^{\left(\perp\right)}_{\gamma\delta}\left(\boldsymbol{q}\right)
\mathcal{G}^{\left(4\right)}_{\varLambda,\alpha\beta\gamma\delta}\left(\omega,\boldsymbol{k},
\varkappa,\boldsymbol{q}\right).\label{3frgstt24}
\end{eqnarray}
Equation (\ref{3frgstt24}) contains the projected four-particle Green's function. We can rewrite this quantity
in terms of \emph{basis four-particle scalar functions} \cite{MonYag}. Such an expression is very cumbersome,
and we do not present it here. We only note that although there is an expansion in some (for example, scalar)
functions, we still need to solve the $n,n+2$ problem. This is only the process of rewriting some unknown
quantities in terms of other unknown quantities.

The only well-known fact about the scalar functions is dictated by the boundary condition (the classical limit)
$\varLambda\rightarrow \infty$. In this limit, the scalar functions are read from the part of the initial total
action which corresponds to the interaction (with the opposite sign).

To obtain more information about ACGF, in particular, about the four-particle scalar functions, it is necessary
to study the symmetries of the problem. We here note that, in some special cases, symmetries can fix an
\textbf{exact} answer for some family of Green's functions (the so-called case of \emph{integrable} problem).
This is not the case of quantum-field model for the NS equation. But the latter still has symmetries, which
permits obtaining some ansatzes for the four-particle scalar functions, and hence for the closure of equation
(\ref{3frgstt24}). Thus, we shall consider the symmetries of our field-theoretic model.

\subsection{Time-gauged Galilean symmetry}

The quantum-field model for the NS equation has the \emph{time-gauged Galilean symmetry}
\cite{LCBDNW1}--\cite{LCBDNW2}. In this case, an infinitesimal transformation of the independent variables on
which the fields depend and of the fields themselves has the form
\begin{eqnarray}
\boldsymbol{r}\rightarrow\boldsymbol{r}'=\boldsymbol{r}+\delta\boldsymbol{r}\left(t\right),\quad
t\rightarrow t'=t,\nonumber\\
\boldsymbol{v}\rightarrow\boldsymbol{v}'=\boldsymbol{v}-\partial_{t}\delta\boldsymbol{r}
\left(t\right)+\left[\delta\boldsymbol{r}\left(t\right)\partial\right]\boldsymbol{v}.\label{4frgstt1}
\end{eqnarray}

The Galilean symmetry, as any other symmetry, generates the so-called \emph{Ward functional identity}, i.e., a
functional equation for the generating functional of a family of Green's functions. The expansion in the fields
of this equation contains some additional (to the FRG-flow equations) information about the corresponding
family of functions. A remarkable property of the Ward functional identity is the fact that it does not contain
the $n,n+2$ problem. Conversely, the expansion in the fields of this equation permits reconstructing the
hierarchy of Green's functions, and this reconstruction starts from the minimal order. The Ward functional
identity therefore \textbf{partially} (for the NS problem) fixes the frequency and the momentum dependencies,
for example, of connected Green's functions.

Strictly speaking, the time-gauged Galilean symmetry is not a symmetry transform\-ation. By definition, the
\emph{symmetry transformation} is an invertible linear transform\-ation of superindices (of all independent
variables on which the fields depend) and the fields with an additional property that \emph{the system action
and the functional integral measure are invariant under this transformation}.

In the case of the Galilean symmetry transformation, the field-theoretic action of the NS equation acquires an
increment \textbf{linear in the fields}. But in this case, there is a simple generalization of the Ward
functional identity which is also free from the $n,n+2$ problem. This generalization, which is also called the
Ward functional identity (we shall omit the word "generalized"$\!$), written for the generating functional of
the connected Green's functions $\mathcal{G}_{c,\varLambda}$, has the form
\begin{eqnarray}
\int_{\boldsymbol{r}}\left\{\delta_{\alpha\beta}\partial_{t}+\left[\partial_{\alpha}\frac{\delta
\mathcal{G}_{c,\varLambda}\left[\,\boldsymbol{j}\right]}{\delta j_{\beta}\left(t,\boldsymbol{r}\right)}
\right]\right\}j_{\beta}\left(t,\boldsymbol{r}\right)=
\int_{\boldsymbol{r}}Q_{\varLambda,\alpha\beta}\left(\partial_{t},\partial\right)\partial_{t}
\frac{\delta\mathcal{G}_{c,\varLambda}\left[\,\boldsymbol{j}\right]}{\delta j_{\beta}
\left(t,\boldsymbol{r}\right)}.\label{4frgstt2}
\end{eqnarray}
In equation (\ref{4frgstt2}), we introduce the compact notation
\begin{eqnarray}
Q_{\varLambda,\alpha\beta}\left(\partial_{t},\partial\right)=N_{\varepsilon,\alpha\beta}
\left(\partial\right)+N_{\alpha\beta}\left(\partial\right)\big(-\partial^{2}_{t}\big)+
R_{\varLambda,\alpha\beta}\left(\partial\right).\label{4frgstt3}
\end{eqnarray}

Our further program is to expand equation (\ref{4frgstt2}) in the fields to obtain relations which express
higher-order connected Green's functions in terms of lower-order connected Green's functions. Such relations
are also Ward identities (without word "functional") for connected Green's functions (in this case). These
identities are more transparent in the $\left(\omega,\boldsymbol{k}\right)$-representation. In this
representation higher-order functions in a \textbf{special} kinematics can be expressed in terms of lower-order
functions. This fact is just the "second player"$\!$ (together with classical limit) who provides information
about the Green's functions and serves as a "support"$\!$ in the construction of the ansatz of the closure of
the flow equation for the two-particle ACGF (\ref{3frgstt24}).

At the same time, to construct such an ansatz is not a goal in this paper. This field of research requires
further independent study. But the theory developed above must suggest an idea of the structure of the
"direct"$\!$ quantum-field approach to the NS problem. The word "direct"$\!$ means that the FRG-flow hierarchy
is solved directly. The main goal in this paper is that we try to find such a solution "in a different
way"$\!$. For this reason, the greatest part of the paper deals with the \emph{holographic approach}, which is
considered in detail in the subsequent section. But before we leave the CFT side and go into the curved space
with additional dimensions, we briefly summarize the results discussed above. The Gaussian part of the action
in the NS field theory can be written as
\begin{eqnarray}
S_{0,\varLambda}\left[\boldsymbol{v}\right]=\int_{t,\boldsymbol{r}}
\mathcal{L}_{0,\varLambda}\left(t,\boldsymbol{r}\right),\quad
\mathcal{L}_{0,\varLambda}\left(t,\boldsymbol{r}\right)=
\frac{1}{2}v_{\alpha}\left(t,\boldsymbol{r}\right)
K_{\varLambda,\alpha\beta}\left(\partial_{t},\partial_{\boldsymbol{r}}\right)
v_{\beta}\left(t,\boldsymbol{r}\right).\label{2hrgtt1}
\end{eqnarray}
The matrix-differential operation $K$ contained in the Lagrangian $\mathcal{L}$ in (\ref{2hrgtt1}) can
standardly be represented as the decomposition into scalar operators
\begin{eqnarray}
K_{\varLambda,\alpha\beta}\left(\partial_{t},\partial_{\boldsymbol{r}}\right)
=K^{\left(\parallel\right)}_{\varLambda}\left(\partial_{t},\partial_{\boldsymbol{r}}\right)
\varPi^{\left(\parallel\right)}_{\alpha\beta}\left(\partial_{\boldsymbol{r}}\right)+
K^{\left(\perp\right)}_{\varLambda}\left(\partial_{t},\partial_{\boldsymbol{r}}\right)
\varPi^{\left(\perp\right)}_{\alpha\beta}\left(\partial_{\boldsymbol{r}}\right).\label{2hrgtt2}
\end{eqnarray}
The explicit form of some elements in formulas (\ref{2hrgtt1})--(\ref{2hrgtt2}) has already been described
above. We also note that, in the last two expressions, we separate the differentiation with respect to
$\boldsymbol{r}$ to make the further expressions in the curved ambient space more convenient.

In the theory under study, we obtained the flow equation (\ref{3frgstt24}) for the two-particle ACGF. To
improve the understanding of the status of this structure and of how to work with it beyond the framework of
different approximations for the four-particle ACGF, it is necessary to consider a simplified ("toy"$\!$)
model. One of such models is based on the (functional) Hamilton--Jacobi equation in the holographic
renormalization group formalism (a "scalar"$\!$ example can be found in \cite{LMPV}), and we pass to the
description of this model.

\section{Hydrodynamic equations\\ from the AdS side}\label{p3}

\subsection{Functional Hamilton--Jacobi equation}

Up to this point, we considered the Navier--Stokes field theory which "lives"$\!$ in the real (physical)
space-time of dimension $D+1$. Now we pose the question of what happens if the field theory with the same
Lagrangian (action) is formulated in the curved space (for brevity, we omit the word "time"$\!$) of a greater
dimension $D+1+p$. In this case, in the framework of statistical physics, we assume that the signature of such
a space is everywhere \textbf{positive}. We also consider only the case of \textbf{diagonal} metric tensors
whose components depend only on the coordinates \textbf{additional} to $\left(t,\boldsymbol{r}\right)$. In the
simplest case $p=1$, there is one additional coordinate $z$, and the space curvature is a constant variable. An
example of such a space is the AdS space \cite{Natsuume}.

We note that the hypothesis of the AdS/CFT correspondence is formulated at the level of correlation functions
rather than at the Lagrangian or action level. It may happen that the latter do not resemble each other at all.
At the same time, the theories with "similar"$\!$ symmetries and sets of fields can give "similar"$\!$ answers
for some correlation functions. We quote the word "similar"$\!$ on purpose; namely, to reveal its meaning is a
nontrivial practical problem. Therefore, it is interesting to study the Hamilton--Jacobi evolution of the NS
field theory embedded in the curved space of dimension $D+1+p$. In this case, as will be shown below, it
suffices first to consider the HJ evolution for the Lagrangian (\ref{2hrgtt1}). Let us formulate its analog in
the curved space for $p=1$. A preliminary change for the velocity field $\boldsymbol{v}$ and the operator $K$
reads
\begin{eqnarray}
v_{\alpha}\left(t,\boldsymbol{r}\right)\rightarrow
v_{\alpha}\left(a\right),\quad
K_{\varLambda,\alpha\beta}\left(\partial_{t},\partial_{\boldsymbol{r}}\right)\rightarrow
K_{\varLambda,z,\alpha\beta}\left(\partial_{a}\right).\label{2hrgtt5}
\end{eqnarray}
Here and hereafter, the quantity $a=\left(t,\boldsymbol{r},z\right)$ denotes a point in the curved (for
example, AdS-coordinate) space. Similarly,
$\partial_{a}=\left(\partial_{t},\partial_{\boldsymbol{r}},\partial_{z}\right)$ denotes differentiation with
respect to all components of~$a$. We sometimes use the notation~$x$ for the pair
$\left(t,\boldsymbol{r}\right)$ (and $\partial_{x}$ for the corresponding differentiation). The quantity
$\varLambda$ again appears in expression (\ref{2hrgtt5}). In the framework of HRG, this quantity denotes all
parameters of the flow without specifying them at the moment.

To write the matrix-differential operation $K$ explicitly, it is necessary to consider the time-gauged Galilean
symmetry. In this case, if the symmetry transformation remains unchanged, i.e., in the form (\ref{4frgstt1}),
then the first specification of $K$ related to the (partial) decomposition into derivatives becomes
\begin{eqnarray}
K_{\varLambda,z,\alpha\beta}\left(\partial_{a}\right)=
K^{\left(0,0\right)}_{\varLambda,z,\alpha\beta}\left(\partial_{\boldsymbol{r}}\right)+
K^{\left(1,0\right)}_{\varLambda,z,\alpha\beta}\left(\partial_{\boldsymbol{r}}\right)
\left(-\partial^{2}_{t}\right)+
K^{\left(0,1\right)}_{\varLambda,z,\alpha\beta}\left(\partial_{\boldsymbol{r}}\right)
\left(-\partial^{2}_{z}\right).\label{2hrgtt6}
\end{eqnarray}
Here we note that the matrix $K^{\left(1,0\right)}$ is transverse. We also introduce the notation
$\dot{v}_{\alpha}\left(a\right)=\partial_{z}v_{\alpha}\left(a\right)$ for the derivative with respect to the
\textbf{zero} coordinate $z$ (for convenience, the coordinate numeration in expressions such as the
energy-momentum tensor $T$ begins from zero). After some modification of the matrix $K^{\left(0,1\right)}$,
which is related to the transition of the derivative with respect to $z$ between the velocity fields (the term
$\mathcal{K}$), the Lagrangian of the NS HRG-field theory becomes
\begin{eqnarray}
\mathcal{L}_{0,\varLambda}\left(a\right)=\mathcal{K}_{\varLambda}\left(a\right)+
\mathcal{U}_{\varLambda}\left(a\right),\quad
\mathcal{K}_{\varLambda}\left(a\right)=\frac{1}{2}\dot{v}_{\alpha}\left(a\right)
K^{\left(0,1\right)}_{\varLambda,z,\alpha\beta}\left(\partial_{\boldsymbol{r}}\right)
\dot{v}_{\beta}\left(a\right),\nonumber\\
\mathcal{U}_{\varLambda}\left(a\right)=\frac{1}{2}v_{\alpha}\left(a\right)
\left[K^{\left(0,0\right)}_{\varLambda,z,\alpha\beta}\left(\partial_{\boldsymbol{r}}\right)+
K^{\left(1,0\right)}_{\varLambda,z,\alpha\beta}\left(\partial_{\boldsymbol{r}}\right)
\left(-\partial^{2}_{t}\right)\right]v_{\beta}\left(a\right).\label{2hrgtt7}
\end{eqnarray}
The Lagrangian (\ref{2hrgtt7}) underlies the whole further HRG-flow procedure.

Further, we reconstruct the energy-momentum tensor $T$ and the (canonically) conjugate momentum field
$\boldsymbol{w}$ from the Lagrangian (\ref{2hrgtt7}). If the Lagrangian (\ref{2hrgtt7}) is independent of the
derivatives of the metric tensor, then the expressions for $T$ and $\boldsymbol{w}$ are
\begin{eqnarray}
T_{ij}=2\frac{\partial\mathcal{L}}{\partial g^{ij}}-g_{ij}\mathcal{L},\quad
\boldsymbol{w}_{i}=\frac{\partial\mathcal{L}}{\partial^{i}\boldsymbol{v}}.\label{2hrgtt8}
\end{eqnarray}
In what follows, it is interesting to consider the zero ($z$-) components of $T$ and $\boldsymbol{w}$. We
denote $g_{00}=g_{z}$ ($g^{00}=1/g_{z}$ because the metric tensor is diagonal) and obtain the expressions
\begin{eqnarray}
T_{\varLambda,00}\left(a\right)=g_{z}\left[\mathcal{K}_{\varLambda}\left(a\right)-
\mathcal{U}_{\varLambda}\left(a\right)\right]=g_{z}\mathcal{H}_{\varLambda}\left(a\right),\quad
T^{00}_{\varLambda}\left(a\right)=\frac{1}{g_{z}}\mathcal{H}_{\varLambda}\left(a\right).\label{2hrgtt9}
\end{eqnarray}
\begin{eqnarray}
w_{0,\alpha}\left(a\right)=g_{z}K^{\left(0,1\right)}_{\varLambda,z,\alpha\beta}
\left(\partial_{\boldsymbol{r}}\right)\dot{v}_{\beta}\left(a\right),\quad
w^{0}_{\alpha}\left(a\right)=K^{\left(0,1\right)}_{\varLambda,z,\alpha\beta}
\left(\partial_{\boldsymbol{r}}\right)\dot{v}_{\beta}\left(a\right).\label{2hrgtt10}
\end{eqnarray}
In expressions (\ref{2hrgtt9})--(\ref{2hrgtt10}), the covariant (subscripts) and contravariant (superscripts)
quantities are distinguished. But the further ambiguity in the construction of the HRG-flow procedure makes
this difference insignificant. This ambiguity seems to be removed by the requirements that some properties be
satisfied in the final differential equations for some Green's function.

The expression for $\boldsymbol{w}_{0}$ must be substituted into $T_{00}$. Then the further change immediately
follows:
\begin{eqnarray}
w_{0,\alpha}\left(a\right)\rightarrow\frac{f\left(z\right)}{\sqrt{|g\left(z\right)|}}\,
\mathcal{G}^{\left(1\right)}_{\varLambda,\alpha}\left[\boldsymbol{v}\right]\left(a\right),\quad
\mathcal{G}^{\left(1\right)}_{\varLambda,\alpha}\left[\boldsymbol{v}\right]\left(a\right)=
\frac{\delta\mathcal{G}_{\varLambda}\left[\boldsymbol{v}\right]}
{\delta v_{\alpha}\left(a\right)}.\label{2hrgtt11}
\end{eqnarray}
In other words, the quantity $\boldsymbol{w}_{0}$ must be expressed in terms of the first functional derivative
of some generating functional $\mathcal{G}$ (we use the same notation as for the ACGF generating functional in
the FRG method, but what we mean is always clear from the context), and the functional $\mathcal{G}$ is called
the \emph{boundary action}. Let us explain this definition.

In the standard approach, the functional $\mathcal{G}$ is defined on the field configurations $\boldsymbol{v}$
which "live"$\!$ not in the entire $D+1+p$-dimensional curved space but only on its $D+1$-dimensional boundary
(the values additional to the $\left(t,\boldsymbol{r}\right)$ coordinate are assumed to be some prescribed
parameters) \cite{IHJP}--\cite{VBMGAL}. An example of the boundary in the AdS space is the domain $z=\ell$
($\ell$ can either be finite or tend to zero thus corresponding to some statement of the problem). All this
explains the meaning of the name of $\mathcal{G}$. In the present paper, we adhere to a different ideology for
the generating functional $\mathcal{G}$; namely, we assume that $\mathcal{G}$ is defined on the field
configurations $\boldsymbol{v}$ which "live"$\!$ in a \textbf{layer} of the $D+1+p$-dimensional curved space.
An example of such a layer in the AdS space can be a domain $a<z<b$ ($a$ and $b$ take various values depending
on the statement of the problem). And the name "boundary action"$\!$ is preserved in this paper.

Such a definition can be justified by our desire to formulate the Hamilton--Jacobi dynamics, which is an
evolution equation with the first derivative with respect to some RG-time in the left-hand side. The right-hand
side contains the squared first (functional) derivative of the boundary action $\mathcal{G}$ with respect to
the argument $\boldsymbol{v}$. At first sight, in such a structure, it suffices to define the functional
$\mathcal{G}$ only for the field variable $\boldsymbol{v}$ which takes values on a certain boundary consistent
with the RG-time. For example, if the role of the RG-time is played by $\ell$, it suffices to consider the
values of $\boldsymbol{v}$ for $z=\ell$. At the same time, the HJ equation can be rewritten in integral form
after integration with respect to the RG-time within some limits. In such a structure, $\mathcal{G}$ must
already be defined for the values of $\boldsymbol{v}$ in a layer consistent with the RG-time interval over
which the integral was calculated. For example, if the RG-time interval is $\left(a,b\right)$, then
$\boldsymbol{v}$ must also be taken into account for all values of $z$ in this interval.

The fact that expression (\ref{2hrgtt11}) contains a function $f$ is also important. This function is an
example of a separate degree of freedom in the construction of the HRG-flow procedure. And this ambiguity is
not a drawback of the method in the case where some "natural"$\!$ conditions for the choice of $f$ can be
formulated. We illustrate such conditions and their consequences below with an example of the equation for the
two-point BGF.

The generating functional $\mathcal{G}$ produces its own family of Green's functions, which we call a boundary
family. In this paper, we first obtain an analytic answer for the two-point BGF and then discuss the
relationship between the boundary Green's functions (BGF) and some family of Green's functions "related to"$\!$
CFT. We now proceed further and define the integration in the HRG method. The transition from the integration
over the volume in CFT to similar integration in the curved space reads
\begin{eqnarray}
\int_{t,\boldsymbol{r}}\rightarrow\int_{a}=\int dt\int d^{D}r\int dz
\sqrt{|g\left(z\right)|}.\label{2hrgtt12}
\end{eqnarray}
The quantity $g\left(z\right)$ in expression (\ref{2hrgtt12}) is the determinant of a metric tensor. To write
the HJ equation, we need the notion of integration over the boundary $\partial a$ of the curved space. The
transition from the integration over the volume to the integration over $\partial a$ can be written as
\begin{eqnarray}
\int_{a}\rightarrow\int_{\partial a}=\int dt\int d^{D}r\sqrt{\frac{|g\left(z\right)|}
{g_{z}\left(z\right)}}.\label{2hrgtt13}
\end{eqnarray}

The last step in the derivation of the functional HJ equation is the following change of functions (such a
change permits formulating the final equation more compactly):
\begin{eqnarray}
f_{g}\left(z\right)=\frac{f\left(z\right)}{g_{z}\left(z\right)
\sqrt{|g\left(z\right)|}},\quad\varGamma_{g}\left(z\right)=
\frac{\varGamma\left(z\right)}{\sqrt{g_{z}\left(z\right)|g\left(z\right)|}}.\label{2hrgtt14}
\end{eqnarray}
In addition to the function $f$, expression (\ref{2hrgtt14}) contains a function $\varGamma$. This function is
a certain "kinetic coefficient"$\!$ in the sense that it is at the derivative with respect to the flow
variable~$z$ in the left-hand side of the HJ equation:
\begin{eqnarray}
\mathrm{lhs\,HJ}=-\varGamma_{g}\left(z\right)\partial_{z}\mathcal{G}_{\varLambda\left(z\right)}
\left[\boldsymbol{v}\right].\label{2hrgtt15}
\end{eqnarray}
The right-hand side of the desired functional equation reads
\begin{eqnarray}
\mathrm{rhs\,HJ}=\frac{1}{2}\int dt\int d^{D}r\Big\{f_{g}^{2}\left(z\right)
\mathcal{G}^{\left(1\right)}_{\varLambda\left(z\right),\alpha}\left[\boldsymbol{v}\right]
\left(a\right)K^{\left(0,1\right)\left(-1\right)}_{\varLambda\left(z\right),\alpha\beta}
\left(\partial_{\boldsymbol{r}}\right)\mathcal{G}^{\left(1\right)}_{\varLambda\left(z\right),\beta}
\left[\boldsymbol{v}\right]\left(a\right)-\nonumber\\
-v_{\alpha}\left(a\right)\left[K^{\left(0,0\right)}_{\varLambda\left(z\right),\alpha\beta}
\left(\partial_{\boldsymbol{r}}\right)+K^{\left(1,0\right)}_{\varLambda\left(z\right),\alpha\beta}
\left(\partial_{\boldsymbol{r}}\right)\left(-\partial^{2}_{t}\right)\right]
v_{\beta}\left(a\right)\Big\}.\label{2hrgtt16}
\end{eqnarray}
Expressions (\ref{2hrgtt15})--(\ref{2hrgtt16}) represent the functional Hamilton--Jacobi equation~\cite{LMPV}.
This equation is the key point in the HRG approach accepted in this paper. Similarly, the Wilson--Polchinski
functional equation underlies one of the possible FRG versions. But we note that the Wilson--Polchinski
equation is exact, while the Hamilton--Jacobi equation is a certain simplified model.

We also note that the dependence $\varLambda\left(z\right)$ is emphasized in equation
(\ref{2hrgtt15})--(\ref{2hrgtt16}). This dependence is a result of the ambiguity of the coordinate $z$. On the
one hand, different values of $z_{i}$ appear in the equation in terms of the variational derivative with
respect to the velocity field $\boldsymbol{v}$ as an argument of the latter. On the other hand, $z$ is a
parameter of the flow (the RG-time after a certain change of variables). For this reason, it is contained in
$\varLambda$.

In the present paper, we derive and solve only the equation for the two-particle BGF
$\mathcal{G}^{\left(2\right)}$. Such an equation is the first nontrivial equation in the hierarchy generated by
the functional equation (\ref{2hrgtt15})--(\ref{2hrgtt16}). In the next subsection, we derive the equation for
$\mathcal{G}^{\left(2\right)}$ itself, and we solve this equation in subsequent subsections. Now let us realize
this program.

\subsection{Equation for the two-particle Green's function}

As usual, the $n$-particle (in this case, boundary) Green's function is determined in terms of the $n$th-order
functional derivative of the functional $\mathcal{G}$ with respect to the field $\boldsymbol{v}$ on the zero
field configuration. For the two-particle BGF $\mathcal{G}^{\left(2\right)}$, we have the expression
\begin{eqnarray}
\mathcal{G}^{\left(2\right)}_{\varLambda\left(z\right),\alpha_{1}\alpha_{2}}
\left[\boldsymbol{v}\right]\left(a_{1},a_{2}\right)
=\frac{\delta^{2}\mathcal{G}_{\varLambda\left(z\right)}
\left[\boldsymbol{v}\right]}{\delta v_{\alpha_{1}}\!\left(a_{1}\right)\delta
v_{\alpha_{2}}\!\left(a_{2}\right)},\nonumber\\
\mathcal{G}^{\left(2\right)}_{\varLambda\left(z\right),\alpha_{1}\alpha_{2}}
\left[\boldsymbol{v}=\mathbf{0}\right]\left(a_{1},a_{2}\right)=
\mathcal{G}^{\left(2\right)}_{\varLambda\left(z\right),\alpha_{1}\alpha_{2}}
\left(a_{1},a_{2}\right).\label{2hrgtt17}
\end{eqnarray}

As previously noted, the hierarchy of integro-differential equations for the BGF in the HJ theory does not
contain the $n,n+1$ problem. It is easy to verify this fact considering the equation for
$\mathcal{G}^{\left(2\right)}$ as an example. We calculate the second functional derivative of equation
(\ref{2hrgtt15})--(\ref{2hrgtt16}) and then set $\boldsymbol{v}=\mathbf{0}$. As a result, we obtain the desired
equation for $\mathcal{G}^{\left(2\right)}$. The left-hand side of the equation reads
\begin{eqnarray}
\mathrm{lhs}=-\varGamma_{g}\left(z\right)\partial_{z}
\mathcal{G}^{\left(2\right)}_{\varLambda\left(z\right),\alpha_{1}\alpha_{2}}
\left(a_{1},a_{2}\right).\label{2hrgtt18}
\end{eqnarray}
We note that, at the moment, the quantities $z$, $z_{1}$, and $z_{2}$ are independent! For the right-hand side
of the equation, we have
\begin{eqnarray}
\mathrm{rhs}=\!\!\int d^{D+1}x\Big\{f_{g}^{2}\left(z\right)
\mathcal{G}^{\left(2\right)}_{\varLambda\left(z\right),\alpha_{1}\beta_{1}}
\left(a_{1},a\right)K^{\left(0,1\right)\left(-1\right)}_{\varLambda\left(z\right),
\beta_{1}\beta_{2}}\left(\partial_{\boldsymbol{r}}\right)
\mathcal{G}^{\left(2\right)}_{\varLambda\left(z\right),\beta_{2}\alpha_{2}}
\left(a,a_{2}\right)-\nonumber\\
-\delta^{\left(D+2\right)}\left(a-a_{1}\right)\left[K^{\left(0,0\right)}_{\varLambda
\left(z\right),\alpha_{1}\alpha_{2}}\left(\partial_{\boldsymbol{r}}\right)+
K^{\left(1,0\right)}_{\varLambda\left(z\right),\alpha_{1}\alpha_{2}}
\left(\partial_{\boldsymbol{r}}\right)\left(-\partial^{2}_{t}\right)\right]
\delta^{\left(D+2\right)}\left(a-a_{2}\right)\Big\}.\label{2hrgtt19}
\end{eqnarray}

Now we denote $x_{i}=\left(t_{i},\boldsymbol{r}_{i}\right)$ and seek the solution in the form
\begin{eqnarray}
\mathcal{G}^{\left(2\right)}_{\varLambda\left(z\right),\alpha_{1}\alpha_{2}}
\left(a_{1},a_{2}\right)=\mathcal{G}^{\left(2\right)}_{\varLambda
\left(z\right),\alpha_{1}\alpha_{2}}\left(z_{i};x_{1},x_{2}\right)
\delta\left(z_{1}-z_{2}\right).\label{2hrgtt20}
\end{eqnarray}
Here we also note that the reduced function in the right-hand side of expression (\ref{2hrgtt20}) depends on
$z_{i}$ ($i=1$ or $2$, which is of no importance because of the Dirac $\delta$ function)! We substitute the
(exact) ansatz (\ref{2hrgtt20}) into equation (\ref{2hrgtt18})--(\ref{2hrgtt19}), set $z_{i}=z$, renormalize
the kinetic coefficient $\varGamma_{g}$ with respect to the value $\delta\left(0\right)$ arising if the
equality $z_{i}=z$ holds, and preserve its notation. The left-hand side of the equation for the (reduced, as we
always mean below) function $\mathcal{G}^{\left(2\right)}$ becomes rather interesting:
\begin{eqnarray}
\mathrm{lhs}=-\varGamma_{g}\left(z\right)\left[\partial_{z}
\mathcal{G}^{\left(2\right)}_{\varLambda\left(z\right),\alpha_{1}\alpha_{2}}
\left(z_{i};x_{1},x_{2}\right)\right]\Big|_{z_{i}=z}.\label{2hrgtt21}
\end{eqnarray}
It follows from expression (\ref{2hrgtt21}) that, in the left-hand side of the flow equation, one can now
obtain a kinetic coefficient depending on the dynamic variables, for example, the momentum and the frequency.
This important degree of freedom is used in subsequent subsections of the paper. Now let us consider the
right-hand side of the equation for~$\mathcal{G}^{\left(2\right)}$:
\begin{eqnarray}
\mathrm{rhs}=\!\!\int d^{D+1}x\Big\{f_{g}^{2}\left(z\right)
\mathcal{G}^{\left(2\right)}_{\varLambda\left(z\right),\alpha_{1}\beta_{1}}
\left(z;x_{1},x\right)K^{\left(0,1\right)\left(-1\right)}_{\varLambda\left(z\right),
\beta_{1}\beta_{2}}\left(\partial_{\boldsymbol{r}}\right)
\mathcal{G}^{\left(2\right)}_{\varLambda\left(z\right),\beta_{2}\alpha_{2}}
\left(z;x,x_{2}\right)-\nonumber\\
-\delta^{\left(D+1\right)}\left(x-x_{1}\right)\left[K^{\left(0,0\right)}_{\varLambda
\left(z\right),\alpha_{1}\alpha_{2}}\left(\partial_{\boldsymbol{r}}\right)+
K^{\left(1,0\right)}_{\varLambda\left(z\right),\alpha_{1}\alpha_{2}}
\left(\partial_{\boldsymbol{r}}\right)\left(-\partial^{2}_{t}\right)\right]
\delta^{\left(D+1\right)}\left(x-x_{2}\right)\Big\}.\label{2hrgtt22}
\end{eqnarray}
It follows from (\ref{2hrgtt22}) that if the system is translation invariant with respect to the coordinates
$\left(t,\boldsymbol{r}\right)$, then such a relation becomes algebraic in the momentum-frequency
representation. Therefore, we should rewrite equation (\ref{2hrgtt21})--(\ref{2hrgtt22}) in the
$\left(\omega,\boldsymbol{k}\right)$-representation.

\subsection{Momentum-frequency representation}

The momentum-frequency representation is "natural"$\!$, because all expressions are have the simplest and most
visual form in this representation. It is easy to show that the integro-differential equation for
$\mathcal{G}^{\left(2\right)}$ becomes only differential in the
$\left(\omega,\boldsymbol{k}\right)$-representation. The left-hand side of this equation reads
\begin{eqnarray}
\mathrm{lhs}=-\varGamma_{g}\left(z\right)\left[\partial_{z}
\mathcal{G}_{\varLambda\left(z\right),\alpha_{1}\alpha_{2}}
\left(z_{i};\omega,\boldsymbol{k}\right)\right]\Big|_{z_{i}=z}.\label{2hrgtt23}
\end{eqnarray}
In the right-hand side, we have the expression
\begin{eqnarray}
\mathrm{rhs}=f_{g}^{2}\left(z\right)
\mathcal{G}_{\varLambda\left(z\right),\alpha_{1}\beta_{1}}
\left(z;-\omega,-\boldsymbol{k}\right)K^{\left(0,1\right)
\left(-1\right)}_{\varLambda\left(z\right),\beta_{1}\beta_{2}}\left(\boldsymbol{k}\right)
\mathcal{G}_{\varLambda\left(z\right),\beta_{2}\alpha_{2}}
\left(z;\omega,\boldsymbol{k}\right)-\nonumber\\
-\left[K^{\left(0,0\right)}_{\varLambda\left(z\right),\alpha_{1}\alpha_{2}}
\left(\boldsymbol{k}\right)+K^{\left(1,0\right)}_{\varLambda\left(z\right),\alpha_{1}\alpha_{2}}
\left(\boldsymbol{k}\right)\omega^{2}\right],\label{2hrgtt24}
\end{eqnarray}
which is quadratic in $\mathcal{G}^{\left(2\right)}$.

Equation (\ref{2hrgtt23})--(\ref{2hrgtt24}) is still very complicated due to the matrix structure with respect
to the indices $\alpha_{i}$ and $\beta_{j}$. To obtain scalar expressions, we project this flow equation on the
longitudinal and transverse directions. First, we decompose the function $\mathcal{G}^{\left(2\right)}$ into
projection operators:
\begin{eqnarray}
\mathcal{G}_{\varLambda\left(z\right),\alpha_{1}\alpha_{2}}
\left(z_{i};\omega,\boldsymbol{k}\right)=
\mathcal{G}^{\left(\parallel\right)}_{\varLambda\left(z\right)}
\left(z_{i};\omega,\boldsymbol{k}\right)
\varPi^{\left(\parallel\right)}_{\alpha_{1}\alpha_{2}}\left(\boldsymbol{k}\right)+
\mathcal{G}^{\left(\perp\right)}_{\varLambda\left(z\right)}
\left(z_{i};\omega,\boldsymbol{k}\right)\varPi^{\left(\perp\right)}_{\alpha_{1}\alpha_{2}}
\left(\boldsymbol{k}\right).\label{2hrgtt25}
\end{eqnarray}
We assume that the scalar functions $\mathcal{G}^{\left(\sigma\right)}$ (the index $\sigma$ denotes the
longitudinal and transverse directions) contained in expression (\ref{2hrgtt25}) are even in the
frequency~$\omega$ and depend only on the modulus of the momentum~$k$. In this case, the left-hand side of the
equation for $\mathcal{G}^{\left(\sigma\right)}$ reads
\begin{eqnarray}
\mathrm{lhs}=-\varGamma_{g}\left(z\right)\left[\partial_{z}
\mathcal{G}^{\left(\sigma\right)}_{\varLambda\left(z\right)}
\left(z_{i};\omega,k\right)\right]\Big|_{z_{i}=z}.\label{2hrgtt26}
\end{eqnarray}
In the right-hand side of the equation for $\mathcal{G}^{\left(\sigma\right)}$, we have
\begin{eqnarray}
\mathrm{rhs}=\frac{f_{g}^{2}\left(z\right)}{K^{\left(\sigma;0,1\right)}_{\varLambda
\left(z\right)}\left(k\right)}\left[\mathcal{G}^{\left(\sigma\right)}_{\varLambda
\left(z\right)}\left(z;\omega,k\right)\right]^{2}-\left[K^{\left(\sigma;0,0\right)}_{\varLambda
\left(z\right)}\left(k\right)+K^{\left(\sigma;1,0\right)}_{\varLambda\left(z\right)}
\left(k\right)\omega^{2}\right].\label{2hrgtt27}
\end{eqnarray}
Equation (\ref{2hrgtt26})--(\ref{2hrgtt27}) practically coincides with the Riccati equation. But there is also
an important difference related to the substitution $z_{i}=z$ participating in (\ref{2hrgtt26}). This
(multivalued) moment will be determined below.

In what follows, we consider only the transverse flow. First, we introduce the notation
\begin{eqnarray}
\mathcal{G}^{\left(\perp\right)}_{\varLambda\left(z\right)}\left(z;\omega,k\right)=
\mathcal{G}_{\varLambda\left(z\right)}\left(z;\omega,k\right),\quad
K^{\left(\perp;0,1\right)}_{\varLambda\left(z\right)}\left(k\right)=
K_{\varLambda\left(z\right)}\left(k\right).\label{2hrgtt28}
\end{eqnarray}
Now we introduce the squared diffusion operator in the $\left(\omega,\boldsymbol{k}\right)$-representation:
\begin{eqnarray}
L^{T}L\left(z;\omega,k\right)=\frac{1}{\bar{g}_{t}}\,\omega^{2}+
\frac{1}{\bar{g}_{\boldsymbol{r}}^{2}}\,\nu_{g}^{2}\left(z\right)k^{4}.\label{2hrgtt29}
\end{eqnarray}
Expression (\ref{2hrgtt29}) contains some elements of a metric tensor. These elements are determined in the
following subsection. We rewrite the functions $K^{\left(0,0\right)}$ and $K^{\left(1,0\right)}$ in the new
notation as follows:
\begin{eqnarray}
K^{\left(0,0\right)}_{\varLambda\left(z\right)}\left(k\right)=
N_{\varLambda\left(z\right)}\left(k\right)\nu_{g}^{2}\left(z\right)k^{4}+
\varDelta N_{\varLambda\left(z\right)}\left(k\right),\quad
K^{\left(1,0\right)}_{\varLambda\left(z\right)}\left(k\right)=
N_{\varLambda\left(z\right)}\left(k\right).\label{2hrgtt30}
\end{eqnarray}

Finally, the left-hand side of the transverse flow equation for the two-particle BGF reads:
\begin{eqnarray}
\mathrm{lhs}=-\varGamma_{g}\left(z\right)\left[\partial_{z}\mathcal{G}_{\varLambda\left(z\right)}
\left(z_{i};\omega,k\right)\right]\Big|_{z_{i}=z}.\label{2hrgtt31}
\end{eqnarray}
In the right-hand side of the flow equation for the two-particle BGF, we have
\begin{eqnarray}
\mathrm{rhs}=\frac{f_{g}^{2}\left(z\right)}{K_{\varLambda\left(z\right)}\left(k\right)}
\left[\mathcal{G}_{\varLambda\left(z\right)}\left(z;\omega,k\right)\right]^{2}-
\boxed{N_{\varLambda\left(z\right)}\left(k\right)L^{T}L\left(z;\omega,k\right)}-
\varDelta N_{\varLambda\left(z\right)}\left(k\right).\label{2hrgtt32}
\end{eqnarray}
The framed expression is the classical Lagrangian (\ref{2hrgtt1}) that is "embedded"$\!$ in the curved space
with one additional coordinate $z$, for example, in the anti-de Sitter space. Because differential equation
(\ref{2hrgtt31})--(\ref{2hrgtt32}) are inhomogeneous, the Lagrangian (\ref{2hrgtt1}) "launches"$\!$ a
holographic flow.

For applications we note that, if equation (\ref{2hrgtt31})--(\ref{2hrgtt32}) is reduced to the Riccati
equation, then the latter can be transformed into a \emph{second}-order \emph{linear} equation, which is
sometimes more convenient.

Equation (\ref{2hrgtt31})--(\ref{2hrgtt32}) still contains a giant ambiguity, for example, in the form of the
function $K$ participating in expression (\ref{2hrgtt28}) or in the form of the metric present in expression
(\ref{2hrgtt29}). To remove this ambiguity, we consider the well-known methods for constructing minimal
holographic models.

\subsection{Minimal model}

To develop a certain intuition about the construction of various holographic model, we recall the foundations
of the general theory of relativity (GTR). It is well known that the general form of the interaction (action)
between the scalar field $\varphi$ and the gravity reads
\begin{eqnarray}
S_{int}\left[\varphi,R\right]=\int_{x}\mathcal{L}_{int}\left[\varphi,R\right]
\left(x\right),\quad\mathcal{L}_{int}\left[\varphi,R\right]=
\sum\limits_{n=0}^{\infty}\frac{1}{n!}U_{n}\left(\varphi\right)R^{n}.\label{minmod1}
\end{eqnarray}
The quantity $R$ contained in expression (\ref{minmod1}) is called the space curvature scalar. This quantity is
the convolution of the Ricci curvature tensor $R_{ij}$ with respect to its two indices. In turn, the tensor
$R_{ij}$ is the convolution of the Riemann curvature tensor $R_{nimj}$ with respect to the indices $n$ and $m$.
The tensor $R_{nimj}$ is a basic characteristic of the space geometry. In connection with expression
(\ref{minmod1}), we also note that the \emph{analyticity assumption} is here satisfied automatically (there are
no fractional powers and no fractional derivatives).

Let us consider the so-called \emph{minimal interaction}:
\begin{eqnarray}
\int_{x}=\int d^{D}x\sqrt{|g\left(x\right)|},\quad
\mathcal{L}_{int}\left[\varphi,R\right]\rightarrow
\mathcal{L}_{int}\left[\varphi,0\right].\label{minmod2}
\end{eqnarray}
The point is that this interaction is realized only through the measure of integration. An important property
of this interaction is the fact that it does not contain any additional "elements"$\!$ whose determination
requires some additional \emph{conditions of choice} (the only function $U_{0}$ in the sum (\ref{minmod1}),
which remains in the Lagrangian, is the potential of the field $\varphi$). Another convenient property of the
choice (\ref{minmod2}) is that the definition of the energy-momentum tensor $T$ of the field~$\varphi$ contains
the derivatives only with respect to the metric tensor but not with respect to its own derivatives.

The total action of the scalar field in the minimal model reads
\begin{eqnarray}
S_{sc}\left[\varphi\right]=\int_{x}\mathcal{L}_{sc}\left[\varphi\right]
\left(x\right),\quad\mathcal{L}_{sc}\left[\varphi\right]=\frac{1}{2}
\left(\partial\varphi\right)^{2}-V\left(\varphi\right).\label{minmod3}
\end{eqnarray}
In expression (\ref{minmod3}), we introduce the standard notation for the squared gradient of the field which
is also related to the metric of the space:
\begin{eqnarray}
\left(\partial\varphi\right)^{2}=g^{ij}\left(x\right)\partial_{i}\varphi
\partial_{j}\varphi=\frac{1}{g_{z}}\left(\partial_{z}\varphi\right)^{2}+
\frac{1}{g_{t}}\left(\partial_{t}\varphi\right)^{2}+\frac{1}{g_{\boldsymbol{r}}}
\left(\partial_{\boldsymbol{r}}\varphi\right)^{2}.\label{minmod4}
\end{eqnarray}
As previously noted, we consider only diagonal metrics with positive signature. In this case, the components of
the metric tensor are equal to $g_{\boldsymbol{r}}$ for all spatial coordinates $\boldsymbol{r}$. Such a metric
has a rather simple form
\begin{eqnarray}
ds^{2}=g_{ij}\left(x\right)dx^{i}dx^{j}=g_{z}\left(dz\right)^{2}+g_{t}\left(dt\right)^{2}+
g_{\boldsymbol{r}}\left(d\boldsymbol{r}\right)^{2}.\label{minmod5}
\end{eqnarray}
Finally, we write the canonical dimensions of the quantities contained in the two last expressions
\begin{eqnarray}
\left[z\right]=\left[t\right]=\left[\boldsymbol{r}\right]=E^{-1},\quad
\left[g^{ij}\right]=\left[g_{ij}\right]=1.\label{minmod6}
\end{eqnarray}
We note that, in the GTR, there is an enormous ambiguity in the choice of some coordinates and hence of their
canonical dimensions. In this paper, the coordinates are introduced so that we have as much as possible in
common with the field theory in the flat space.

Everything discussed above is the standard statement of the theory of the real scalar field $\varphi$ in the
presence of gravity. For our further purposes, this material forms some lessons which it is necessary to learn
to construct an HRG model of the stochastic theory of turbulence. The above discussion is not a rigorous proof,
it only suggest how to act. Equipped with this knowledge, we can construct more complicated models, and an
example of such models is given below.

Let us consider the "diffusion"$\!$ Lagrangian. In the plane space, such a Lagrangian can be obtained from
expression (\ref{minmod3}) if, in the latter, we use the flat metric and perform the change
\begin{eqnarray}
\left(\partial\varphi\right)^{2}\rightarrow A\left[\left(\partial_{t}\varphi\right)^{2}+
\xi\left(\partial_{z}\varphi\right)^{2}+\nu^{2}\left(\partial_{\boldsymbol{r}}^{2}
\varphi\right)^{2}\right].\label{minmod7}
\end{eqnarray}
In this case, the canonical dimensions of the quantities are
\begin{eqnarray}
\left[z\right]=\left[t\right]=\left[\boldsymbol{r}\right]^{2}=E^{-2},\quad
\left[\xi\right]=\left[\nu\right]=1.\label{minmod8}
\end{eqnarray}

Now we "switch on"$\!$ the metric. In this case, the canonical dimensions of the metric tensor can be chosen as
\begin{eqnarray}
\left[g_{z}\right]=\left[g_{t}\right]=E^{2},\quad
\left[g_{\boldsymbol{r}}\right]=1.\label{minmod9}
\end{eqnarray}
We note that the following example of a metric is the closest to the classical AdS space:
\begin{eqnarray}
g_{t}=\xi g_{z}=\varLambda^{2}g_{\boldsymbol{r}},\quad
g_{\boldsymbol{r}}=\frac{1}{\left(\varLambda^{2}z\right)^{\alpha}}.\label{minmod10}
\end{eqnarray}
The standard value of $\alpha$ is two, but it can be treated as a parameter of the model. In a more general
case, the diagonal metric of our problem can be represented as
\begin{eqnarray}
g_{z}\left(z\right)=\varLambda^{2}\bar{g}_{z}\left(\varLambda^{2}z\right),\quad
g_{t}\left(z\right)=\varLambda^{2}\bar{g}_{t}\left(\varLambda^{2}z\right),\quad
g_{\boldsymbol{r}}\left(z\right)=\bar{g}_{\boldsymbol{r}}\left(\varLambda^{2}z\right).\label{minmod11}
\end{eqnarray}
We have already encountered the dimensionless functions $\bar{g}_{z}$, $\bar{g}_{t}$, and
$\bar{g}_{\boldsymbol{r}}$ (of the dimensionless argument $\varLambda^{2}z$), which participate in expression
(\ref{minmod11}), in the definition of the squared diffusion operator (\ref{2hrgtt29}).

Now we consider the "diffusion"$\!$ Lagrangian in the curved space, more precisely, the change of the kinetic
term required to obtain such a Lagrangian. In the case of the metric (\ref{minmod10}), this change reads
\begin{eqnarray}
\left(\partial\varphi\right)^{2}\rightarrow A\left\{\frac{1}{g_{\boldsymbol{r}}}
\left[\left(\partial_{t}\varphi\right)^{2}+\xi\left(\partial_{z}\varphi\right)^{2}\right]+
\frac{\nu^{2}}{g_{\boldsymbol{r}}^{2}}\left(\partial_{\boldsymbol{r}}^{2}
\varphi\right)^{2}\right\}.\label{minmod12}
\end{eqnarray}
In the general case, we have
\begin{eqnarray}
\left(\partial\varphi\right)^{2}\rightarrow A\left\{\frac{1}{\bar{g}_{t}}
\left(\partial_{t}\varphi\right)^{2}+\frac{1}{\bar{g}_{z}}\left(\partial_{z}\varphi\right)^{2}+
\frac{\nu^{2}}{\bar{g}_{\boldsymbol{r}}^{2}}\left(\partial_{\boldsymbol{r}}^{2}
\varphi\right)^{2}\right\}.\label{minmod13}
\end{eqnarray}
All above expressions explain how to construct an "intermediate"$\!$ field theory between the "scalar
classics"$\!$ and the stochastic theory of turbulence. In the light of the discussion above, we now turn to the
latter.

As the first step, we place the quantity~$A$ between the fields. Thus, we obtain the standard quadratic form.
After this, we perform the following change (for brevity, we denote the dimensionless argument
$\varLambda^{2}z$ simply by~$z$):
\begin{eqnarray}
A\rightarrow N_{\varLambda\left(z\right),\alpha\beta}\left(\partial_{\boldsymbol{r}}\right)=
N_{0}n_{\alpha\beta}\big(\hat{k}\big),\quad\hat{k}=\frac{\partial_{\boldsymbol{r}}^{2}}
{\bar{g}_{\boldsymbol{r}}\left(z\right)}.\label{minmod14}
\end{eqnarray}
Here $N_{0}$ is a constant. For the functions $K$ contained in decomposition (\ref{2hrgtt6}), we now have the
following expressions in terms of the inverse metric tensor and the (inverse) pumping
operator~(\ref{minmod14}):
\begin{eqnarray}
K^{\left(0,1\right)}_{\varLambda\left(z\right),\alpha\beta}\left(\partial_{\boldsymbol{r}}\right)=
\varLambda^{2}g^{z}\left(z\right)N_{\varLambda\left(z\right),\alpha\beta}
\left(\partial_{\boldsymbol{r}}\right),\nonumber\\
K^{\left(1,0\right)}_{\varLambda\left(z\right),\alpha\beta}\left(\partial_{\boldsymbol{r}}\right)=
\varLambda^{2}g^{t}\left(z\right)N_{\varLambda\left(z\right),\alpha\beta}
\left(\partial_{\boldsymbol{r}}\right),\nonumber\\
K^{\left(0,0\right)}_{\varLambda\left(z\right),\alpha\beta}\left(\partial_{\boldsymbol{r}}\right)=
\left[\nu g^{\boldsymbol{r}}\left(z\right)\right]^{2}N_{\varLambda\left(z\right),\alpha\beta}
\left(\partial_{\boldsymbol{r}}\right)\partial_{\boldsymbol{r}}^{4}.\label{minmod15}
\end{eqnarray}
Finally, in the momentum-frequency representation, for all "construction blocks"$\!$ of the transverse flow for
the two-particle BGF, we have
\begin{eqnarray}
K^{\left(0,1\right)}_{\varLambda\left(z\right)}\left(k\right)=
K_{\varLambda\left(z\right)}\left(k\right)=N_{\varLambda\left(z\right)}\left(k\right),\nonumber\\
K^{\left(1,0\right)}_{\varLambda\left(z\right)}\left(k\right)=\frac{1}{\bar{g}_{t}}
N_{\varLambda\left(z\right)}\left(k\right),\quad
K^{\left(0,0\right)}_{\varLambda\left(z\right)}\left(k\right)=\frac{\nu^{2}}
{\bar{g}_{\boldsymbol{r}}^{2}}N_{\varLambda\left(z\right)}\left(k\right)k^{4}.\label{minmod16}
\end{eqnarray}

Now we write the transverse flow equation in the minimal model. The left-hand side of the flow equation reads
\begin{eqnarray}
\mathrm{lhs}=-\varGamma_{g}\left(z\right)\left[\partial_{z}\mathcal{G}_{\varLambda\left(z\right)}
\left(z_{i};\omega,k\right)\right]\Big|_{z_{i}=z}.\label{minmod17}
\end{eqnarray}
In the right-hand side of the flow equation, we have
\begin{eqnarray}
\mathrm{rhs}=\frac{f_{g}^{2}\left(z\right)}{N_{\varLambda\left(z\right)}\left(k\right)}
\left[\mathcal{G}_{\varLambda\left(z\right)}\left(z;\omega,k\right)\right]^{2}-
N_{\varLambda\left(z\right)}\left(k\right)\left(\frac{\omega^{2}}{\bar{g}_{t}}+
\frac{\nu^{2}k^{4}}{\bar{g}_{\boldsymbol{r}}^{2}}\right).\label{minmod18}
\end{eqnarray}
Comparing the result (\ref{minmod17})--(\ref{minmod18}) with equation (\ref{2hrgtt31})--(\ref{2hrgtt32}), we
make several conclusions. Most of the elements of the flow equation are defined in the minimal model: the
"remainder"$\!$ $\varDelta N$ is absent, the viscosity $\nu$ is a constant quantity, and the function $K$ is
expressed in terms of the (inverse) pumping function.

In the framework of the minimal model (\ref{minmod17})--(\ref{minmod18}), we consider the following scenario.
We assume that the metric satisfies the condition $\bar{g}_{t}=\bar{g}_{\boldsymbol{r}}^{2}$. This choice is
standard in the holographic study of the Lifshitz quantum-field models \cite{Taylor} and the Galilean field
theory \cite{SHJJBSK}. We also assume that all coefficient functions of differential equation
(\ref{minmod17})--(\ref{minmod18}) are \textbf{power} functions of~$z$. Such a scenario can easily be realized,
for example, by using the power metric, the power pumping function, and the power functions $\varGamma_{g}$ and
$f_{g}$. Finally, we assume that the two-particle BGF is independent of $z_{i}$. In this case, for the desired
Green's function, we obtain a simple generalization of a special Riccati equation, and its solution can be
obtained in terms of special functions (the modified Bessel functions $I$ and $K$). An important property of
such a solution is the fact that it depends on the \textbf{sum} $\omega^{2}+\nu^{2}k^{4}$ parametrically. Such
solutions have already been discussed in the framework of the scalar field theory in the literature
\cite{Taylor}--\cite{SHJJBSK}.

Our goal in the present paper is more general; namely, to find a model where the dependence
$\omega^{2}+\nu^{2}k^{4}$ \textbf{splits} and the dependence on the (inverse) pumping function $N$ is
"minimal"$\!$ (the meaning of this word is explained below). The search of such a model is also important for
the following reason: this search permits understanding the nature of the holographic approach more explicitly
and also sharpens the intuition about different HRG-flow procedures. To achieve our goal, we now consider
different methods for constructing extended holographic models.

\subsection{Extended model}

In addition to the minimal model, there are many different beautiful holographic scenarios for different
quantities encountered, for example, in the transverse flow equation (\ref{2hrgtt31})--(\ref{2hrgtt32}) for the
two-particle BGF. We start our consideration from different modifications of the viscosity $\nu$. We have the
chain of generalizations
\begin{eqnarray}
\nu\rightarrow Z_{\nu}\left(z\right)\nu\rightarrow
\bar{\nu}_{\varLambda\left(z\right)}\left(k\right);\nonumber\\
Z_{\nu}\left(0\right)=1,\quad\bar{\nu}_{\varLambda\left(0\right)}\left(k\right)=\nu.\label{extmod1}
\end{eqnarray}
We note that the equalities in the second line in (\ref{extmod1}) are significant but unnecessary, and we do
not consider them in this paper.

We make a generalizing (compared with the preceding subsection) assumption about the form of the two-particle
BGF. We now choose the separable ansatz for this quantity:
\begin{eqnarray}
\mathcal{G}_{\varLambda\left(z\right)}\left(z_{i};\omega,k\right)=\mathcal{A}
\left(z_{i};k\right)\mathcal{G}_{\varLambda\left(z\right)}
\left(\omega,k\right).\label{extmod2}
\end{eqnarray}
We note that expression (\ref{extmod2}) is an exact assumption rather than an approximation. The
amplitude~$\mathcal{A}$ plays an important role; namely, using the latter in the left-hand side of the flow
equation (\ref{2hrgtt31}), we obtain the kinetic coefficient depending on the modulus of the momentum $k$ (this
was already briefly discussed above). Indeed, just as at the end of the preceding subsection, in the framework
of the ansatz (\ref{extmod2}), the left-hand side of the transverse flow equation becomes closed in~$z$:
\begin{eqnarray}
\mathrm{lhs}=-\varGamma_{g}\left(z\right)\mathcal{A}\left(z;k\right)\partial_{z}
\mathcal{G}_{\varLambda\left(z\right)}\left(\omega,k\right).\label{extmod3}
\end{eqnarray}

We assume that the actual variable is a function $\chi_{\varLambda\left(z\right)}\left(k\right)$ (an example of
such a function is considered below). Expression (\ref{extmod3}) then becomes
\begin{eqnarray}
\mathrm{lhs}=-\mathcal{A}\left(z;k\right)\left[\varGamma_{g}\left(z\right)\partial_{z}
\ln\chi_{\varLambda\left(z\right)}\left(k\right)\right]\chi\partial_{\chi}
\mathcal{G}_{\chi}\left(\omega,k\right).\label{extmod4}
\end{eqnarray}
Expression (\ref{extmod4}) depends only on the actual variable~$\chi$ in the case where the amplitude
$\mathcal{A}$ satisfies the relation
\begin{eqnarray}
\mathcal{A}\left(z;k\right)=\frac{\mathcal{A}_{\chi}}{\varGamma_{g}\left(z\right)\partial_{z}
\ln\chi_{\varLambda\left(z\right)}\left(k\right)}.\label{extmod5}
\end{eqnarray}
In what follows, we assume that the amplitude $\mathcal{A}_{\chi}$ is equal to~$1$. This implies the simple
form of  the left-hand side of the flow equation
\begin{eqnarray}
\mathrm{lhs}=-\chi\partial_{\chi}\mathcal{G}_{\chi}\left(\omega,k\right).\label{extmod6}
\end{eqnarray}

Now we consider the right-hand side of the flow equation (\ref{2hrgtt32}). In the framework of separable ansatz
(\ref{extmod2}), we have the expression
\begin{eqnarray}
\mathrm{rhs}=\frac{\left[f_{g}\left(z\right)\mathcal{A}\left(z;k\right)\right]^{2}}
{K_{\varLambda\left(z\right)}\left(k\right)}
\left[\mathcal{G}_{\varLambda\left(z\right)}\left(\omega,k\right)\right]^{2}-
N_{\varLambda\left(z\right)}\left(k\right)\left(\omega^{2}+h_{g}\left(z\right)
\bar{\nu}_{\varLambda\left(z\right)}^{2}\left(k\right)k^{4}\right).\label{extmod7}
\end{eqnarray}
From now on and till the end of the paper, we assume that the components of the metric tensor satisfy the
condition \cite{Taylor}--\cite{SHJJBSK}:
\begin{eqnarray}
h_{g}\left(z\right)=\frac{\bar{g}_{t}\left(z\right)}
{\bar{g}_{\boldsymbol{r}}^{2}\left(z\right)}=1.\label{extmod8}
\end{eqnarray}
We also introduce an intermediate function $M$ as follows:
\begin{eqnarray}
M_{\varLambda\left(z\right)}\left(k\right)=
\frac{K_{\varLambda\left(z\right)}\left(k\right)}
{\left[f_{g}\left(z\right)\mathcal{A}\left(z;k\right)\right]^{2}}.\label{extmod9}
\end{eqnarray}
Now expression (\ref{extmod7}) becomes simpler:
\begin{eqnarray}
\mathrm{rhs}=\frac{1}{M_{\varLambda\left(z\right)}\left(k\right)}
\left[\mathcal{G}_{\varLambda\left(z\right)}\left(\omega,k\right)\right]^{2}-
N_{\varLambda\left(z\right)}\left(k\right)\left(\omega^{2}+
\bar{\nu}_{\varLambda\left(z\right)}^{2}\left(k\right)k^{4}\right).\label{extmod10}
\end{eqnarray}
Just as in the case of the left-hand side of the flow equation, we assume that expression (\ref{extmod10})
depends only on the actual variable~$\chi$:
\begin{eqnarray}
\mathrm{rhs}=\frac{1}{M_{\chi}\left(k\right)}\left[\mathcal{G}_{\chi}
\left(\omega,k\right)\right]^{2}-N_{\chi}\left(k\right)\left(\omega^{2}+
\bar{\nu}_{\chi}^{2}\left(k\right)k^{4}\right).\label{extmod11}
\end{eqnarray}
All the above assumptions imply the transverse flow equation for the two-particle BGF in the form
\begin{eqnarray}
\chi\partial_{\chi}\mathcal{G}_{\chi}\left(\omega,k\right)=-\frac{1}{M_{\chi}\left(k\right)}
\left[\mathcal{G}_{\chi}\left(\omega,k\right)\right]^{2}+
N_{\chi}\left(k\right)\left(\omega^{2}+
\bar{\nu}_{\chi}^{2}\left(k\right)k^{4}\right).\label{extmod12}
\end{eqnarray}
Equation (\ref{extmod12}) is a generalization of the result (\ref{minmod17})--(\ref{minmod18}) obtained in the
framework of the minimal model. At the moment, expression (\ref{extmod12}) is still rather general. The further
simplification of this expression is closely related to the evolution of the function~$M$.

We again introduce the function $\rho$ by the rule
\begin{eqnarray}
\rho_{\chi}\left(k\right)=\frac{N_{\chi}\left(k\right)\bar{\nu}_{\chi}^{2}
\left(k\right)k^{4}}{M_{\chi}\left(k\right)}.\label{extmod13}
\end{eqnarray}
Relation (\ref{extmod13}) is convenient for the further choice; namely, we assume that the function $\rho$ is a
constant. Such a choice is possible because the function $\rho$ is the holographic degree of freedom
corresponding to different versions of the HRG-flow procedure. Our choice cannot influence the form of the
Navier--Stokes Lagrangian embedded in the curved space. An interesting problem is to find other versions of the
HRG-flow equations. In this paper, we only consider the version where $\rho$ is a constant. In this case, the
flow equation reads
\begin{eqnarray}
\chi\partial_{\chi}\mathcal{G}_{\chi}\left(\omega,k\right)=-\frac{\rho}
{N_{\chi}\left(k\right)\bar{\nu}_{\chi}^{2}\left(k\right)k^{4}}
\left[\mathcal{G}_{\chi}\left(\omega,k\right)\right]^{2}+
N_{\chi}\left(k\right)\left(\omega^{2}+
\bar{\nu}_{\chi}^{2}\left(k\right)k^{4}\right).\label{extmod14}
\end{eqnarray}
Equation (\ref{extmod14}) is the main equation in this paper from the holographic standpoint. It is important
to note that, under this choice of the flow, the index of the solution (of special functions) is independent of
the frequency $\omega$ and the momentum~$k$ (this can easily be proved by using the scaling transformation)!
Otherwise, such a solution would require a serious justification. Thus, we have obtained an additional
\textbf{condition of choice}.

Further, we specify how all coefficient functions depend on the actual variable $\chi$. Just as in the case of
minimal model, we deal with the power-law dependence
\begin{eqnarray}
N_{\chi}\left(k\right)=N\left(k\right)\chi^{\eta_{n}},\quad
\bar{\nu}_{\chi}\left(k\right)=\nu\left(k\right)\chi^{\eta_{\nu}}.\label{extmod15}
\end{eqnarray}
In this subsection, we assume that the values of the exponents $\eta_{n}$ and $\eta_{\nu}$ are parameters of
the model. Now the flow equation can be written explicitly
\begin{eqnarray}
\chi\partial_{\chi}\mathcal{G}_{\chi}\left(\omega,k\right)=-\frac{\rho}
{A\left(k\right)\chi^{\eta_{n}+2\eta_{\nu}}}
\left[\mathcal{G}_{\chi}\left(\omega,k\right)\right]^{2}+A\left(k\right)
\chi^{\eta_{n}+2\eta_{\nu}}+B\left(\omega,k\right)\chi^{\eta_{n}}.\label{extmod16}
\end{eqnarray}
In expression (\ref{extmod16}), we use the compact notation
\begin{eqnarray}
A\left(k\right)=N\left(k\right)\nu^{2}\left(k\right)k^{4},\quad
B\left(\omega,k\right)=N\left(k\right)\omega^{2}.\label{extmod17}
\end{eqnarray}

Equation (\ref{extmod16}) is a generalization of a special Riccati equation. This equation has an analytic
solution in the form of a wonderful combination of the modified Bessel functions $I$ and $K$. In this case,
such an analytic solution is possible even for a more general equation obtained from (\ref{extmod16}) by
"elongating"$\!$ the derivative in the left-hand side of (\ref{extmod16}):
\begin{eqnarray}
\chi\partial_{\chi}\mathcal{G}_{\chi}\left(\omega,k\right)\rightarrow
\left(\chi\partial_{\chi}-\eta\right)\mathcal{G}_{\chi}\left(\omega,k\right).\label{extmod18}
\end{eqnarray}
This "elongation"$\!$ will be explained in a separate subsection. But in the following subsection, we write and
analyze the analytic solution of equation (\ref{extmod16}) for the two-particle BGF in detail.

\subsection{Analytic solution of the equation\\ for the two-particle Green's function}

As previously noted, the general solution of equation (\ref{extmod16}) with the elongated derivative is
constructed from the modified Bessel functions $I$ and $K$. As a solution of a first-order differential
equation, this solution also contains the constant of integration $C_{\mathcal{G}}$. Moreover, in the general
case, the quantity $C_{\mathcal{G}}$ is a function of the frequency $\omega$ and the modulus of the momentum
$k$. It is important that the asymptotics of the solution as $\chi\rightarrow\infty$ does not provide any
information about $C_{\mathcal{G}}$. The constant $C_{\mathcal{G}}$ is chosen as $\chi\rightarrow 0$. In this
case, if $C_{\mathcal{G}}=0$, then only the function $I$ remains in the solution, and if
$C_{\mathcal{G}}=\infty$, then only the function $K$ survives. These two values of the constant of integration
are distinguished and are analogs of the two "quantization methods"$\!$ (standard and alternative) in the AdS
space \cite{LMPV}, \cite{IRKEW}. In what follows, we mainly consider the standard quantization method, and
hence the solution of equation (\ref{extmod16}) with the elongated derivative for the value
$C_{\mathcal{G}}=0$. This solution has the form
\begin{eqnarray}
\mathcal{G}_{\chi}\left(\omega,k\right)=\frac{A\left(k\right)}{2\rho}\chi^{\eta_{n}+2\eta_{\nu}}
\left\{\eta-\eta_{n}-2\eta_{\nu}+2\chi\partial_{\chi}\ln I_{-\frac{\xi}{2\eta_{\nu}}}\left[x_{\chi}
\left(\omega,k\right)\right]\right\}.\label{sol1}
\end{eqnarray}
The solution (\ref{sol1}) can also be represented in the form without any logarithmic derivative if we use the
well-known relation
\begin{eqnarray}
\chi\partial_{\chi}\ln I_{-\frac{\xi}{2\eta_{\nu}}}\left[x_{\chi}
\left(\omega,k\right)\right]=
\frac{\xi}{2}-\frac{\eta_{\nu}x_{\chi}\left(\omega,k\right)
I_{-\frac{\xi}{2\eta_{\nu}}+1}\left[x_{\chi}\left(\omega,k\right)\right]}
{I_{-\frac{\xi}{2\eta_{\nu}}}\left[x_{\chi}\left(\omega,k\right)\right]}.\label{sol2}
\end{eqnarray}
In the last two relations, we introduced the notation
\begin{eqnarray}
x_{\chi}\left(\omega,k\right)=\frac{1}{\eta_{\nu}}\sqrt{\frac{\rho B\left(\omega,k\right)}
{A\left(k\right)}}\chi^{-\eta_{\nu}}=\frac{\sqrt{\rho}}{\eta_{\nu}}\frac{|\omega|}
{\bar{\nu}_{\chi}\left(k\right)k^{2}},\nonumber\\
\xi=\sqrt{\left(\eta-\eta_{n}-2\eta_{\nu}\right)^{2}+4\rho}.\label{sol3}
\end{eqnarray}
It follows from the definition of the function $x$ that the solution (\ref{sol1}) is an even function of the
frequency $\omega$, just as this must be. Finally, we transform the amplitude of the solution (\ref{sol1}) in
terms of quantities (\ref{extmod15}) as follows:
\begin{eqnarray}
\frac{A\left(k\right)}{2\rho}\chi^{\eta_{n}+2\eta_{\nu}}=
\frac{N_{\chi}\left(k\right)\bar{\nu}_{\chi}^{2}\left(k\right)k^{4}}{2\rho}.\label{sol4}
\end{eqnarray}
Expressions (\ref{sol1})--(\ref{sol4}) completely describe the solution of equation (\ref{extmod16}).

Now we use the obtained solution for the two-particle BGF to reconstruct the "physical"$\!$ two-particle
Green's function (PGF). The latter "lives"$\!$ on the boundary $z=0$ of the curved space (in this case, no
infrared regularization is required). Let the boundary value of the actual variable be
$\chi_{\varLambda\left(z=0\right)}\left(k\right)=\chi\left(k\right)$. In the simplest case, the function
$\chi\left(k\right)$ is a constant, for example, it is equal to~$1$. At this stage, our description is still
general. We introduce the following notation on the boundary:
\begin{eqnarray}
N_{\chi=\chi\left(k\right)}\left(k\right)=N_{B}\left(k\right),\quad
\bar{\nu}_{\chi=\chi\left(k\right)}\left(k\right)=\bar{\nu}_{B}\left(k\right).\label{sol5}
\end{eqnarray}
\begin{eqnarray}
x_{\chi=\chi\left(k\right)}\left(\omega,k\right)=x\left(\omega,k\right),\quad
\mathcal{G}_{\chi=\chi\left(k\right)}\left(\omega,k\right)=
\mathcal{G}\left(\omega,k\right).\label{sol6}
\end{eqnarray}
In terms of (\ref{sol5})--(\ref{sol6}), the expression for PGF reads
\begin{eqnarray}
\mathcal{G}\left(\omega,k\right)=\frac{N_{B}\left(k\right)\bar{\nu}_{B}^{2}\left(k\right)k^{4}}{2\rho}
\left\{\sigma-\frac{2\eta_{\nu}x\left(\omega,k\right)I_{-\frac{\xi}{2\eta_{\nu}}+1}
\left[x\left(\omega,k\right)\right]}{I_{-\frac{\xi}{2\eta_{\nu}}}
\left[x\left(\omega,k\right)\right]}\right\}.\label{sol7}
\end{eqnarray}
In the last expression, we introduced the compact notation
\begin{eqnarray}
x\left(\omega,k\right)=\frac{\sqrt{\rho}}{\eta_{\nu}}\frac{|\omega|}{\bar{\nu}_{B}
\left(k\right)k^{2}},\quad\sigma=\eta-\eta_{n}-2\eta_{\nu}+\xi.\label{sol8}
\end{eqnarray}
Expressions (\ref{sol5})--(\ref{sol8}) describe the holographic version of PGF. Precisely this quantity is the
observable correlation function of velocity (we pay separate attention to this problem below in the discussion
of the meaning of the boundary action and the corresponding BGF family) and can be compared with the
predictions obtained in other theories of turbulence on the one hand and with the experimental results on the
other hand. Moreover, an analysis of this quantity in the framework of the FRG hierarchy is an important
subject of investigation.

The two-particle PGF (\ref{sol7}) separately depends on the frequency $\omega$ and on the modulus of the
momentum $k$. Another important property of the function (\ref{sol7}) is the fact that it "minimally"$\!$
depends on the details of the pumping function. To demonstrate this, we consider the rules for choosing the
random force correlator. This material also explains in detail how the pumping function "is transferred"$\!$
from the CFT into the curved space of a greater dimension, for example, the AdS space.

\subsection{Choice of a random force correlator}

In the stochastic theory of turbulence, the Fourier transform of the random force correlator is usually called
the pumping function $D\left(k\right)$ \cite{Vasilev}--\cite{AAV}. For this reason, we further discuss the
choice of $D\left(k\right)$. What is known about the last of the most general considerations? The random force
$\boldsymbol{f}$ provides a phenomenological model of stochasticity (which must spontaneously arise in reality
as a consequence of instability of the laminar flow) and, simultaneously, of the energy pumping into the system
due to the interaction with large-scale vortices. The average energy pumping power (the energy quantity per
unit time and per unit mass) $W$ is related to the function $D\left(k\right)$ as follows:
\begin{eqnarray}
W=\frac{D-1}{4}\int_{\boldsymbol{k}}D\left(k\right).\label{Vas1}
\end{eqnarray}
Clearly, expression (\ref{Vas1}) is insufficient for the unique choice of $D\left(k\right)$. The stochastic
theory of turbulence does not contain such a rule for choosing the function $D\left(k\right)$ at all.

In the framework of perturbative RG, it is important that, on the one hand, $D\left(k\right)$ be ultraviolet
(the contribution to the integral (\ref{Vas1}) must be generated by the domain of large momenta
$k\sim\varLambda$, and the asymptotics of $D\left(k\right)$ for large $k$ must be polinomial). On the other
hand, $D\left(k\right)$ must admit transition from the ultraviolet form to the infrared form corresponding to
the actual pumping at which the main contribution to the integral (\ref{Vas1}) is generated by the domain of
small momenta $k\sim m$ (the energy pumping by large-scale vortices). An example of such a decomposition of
$D\left(k\right)$ in the ultraviolet problem is
\begin{eqnarray}
D\left(k\right)=D_{0}\,k^{4-D-2\varepsilon}\left(1+\frac{m^{2}}
{k^{2}}\right)^{-\varepsilon},\quad 0<\varepsilon<2.\label{sep1}
\end{eqnarray}
The parameter $\varepsilon$ contained in expression (\ref{sep1}) is an independent parameter of the model and
is in no way related to the dimension of the space $D$. Its value characterizes the \emph{degree of deviation
from the logarithmic state}. The model becomes logarithmic for $\varepsilon=0$, and the (infrared) pumping
becomes realistic only for $\varepsilon>2$ (as a rule, it is assumed that the realistic value of $\varepsilon$
is equal to~$2$). The pumping is ultraviolet in the domain $0<\varepsilon<2$.

The generalized pumping model has the form
\begin{eqnarray}
D\left(k\right)=D_{0}\,k^{4-D-2\varepsilon}d\left(\frac{m}{k}\right),\quad
d\left(0\right)=1,\quad d\left(\infty\right)=0.\label{sep2}
\end{eqnarray}
In expression (\ref{sep2}), it is assumed that the function $d$ is ambiguous but "sufficiently good"$\!$ (in
particular, analytic in $m^{2}$ near zero). The assumption about the \emph{power-law behavior} of
$D\left(k\right)$ for large $k$ is in fact the only aspect in the perturbative RG-theory of turbulence which is
open to criticism. This aspect can be justified by several arguments which are generally acknowledged nowadays.
We note that, in the exact theory, the energy pumping must be generated by the interaction between the pulsing
and smooth components of the velocity, and therefore, its characteristics must principally be computable for a
certain problem (for example, for the flow in a tube with a prescribed pressure drop at the ends). But there is
still no complete general theory of such a type, and in the framework of the stochastic problem, which is only
a certain simplified phenomenological version of this (hypothetic) rigid theory, a specific choice of the
function $D\left(k\right)$ can be justified only by general considerations and results.

An important advantage of the approaches based on the FRG and HRG methods is the fact that we can initially use
the realistic \textbf{infrared} pumping function $D\left(k\right)$ \cite{LCBDNW1}--\cite{LCBDNW2}. As an
example of $D\left(k\right)$ for the infrared problem, we have
\begin{eqnarray}
D\left(k\right)=D_{0}\,m^{-2\varepsilon}k^{4-D}\left(1+\frac{k^{2}}
{m^{2}}\right)^{-\varepsilon},\quad \varepsilon>2.\label{sep3}
\end{eqnarray}
A generalized model of infrared $D\left(k\right)$ states
\begin{eqnarray}
D\left(k\right)=D_{0}\,m^{-2\varepsilon}k^{4-D}d\left(\frac{k}{m}\right),\quad
d\left(0\right)=1,\quad d\left(\infty\right)=0.\label{sep4}
\end{eqnarray}
It follows from expression (\ref{sep4}) that, as the foundation, we can take the function $D\left(k\right)$
containing the product of two "components"$\!$, i.e., the power-law amplitude and an ambiguous function $d$
(the dimension $D_{0}$ of the parameter does not change in this case):
\begin{eqnarray}
D\left(k\right)=D_{0}\left(\frac{k}{m}\right)^{\eta_{d}}d\left(\frac{k}{m}\right).\label{sep5}
\end{eqnarray}
We assume that the parameter $\eta_{d}$ in expression (\ref{sep5}) is positive.

In what follows, we show that the HRG-flow can be organized so that the two-particle PGF (\ref{sol7}) is
independent of the function $d$. This is precisely the meaning of the "minimal"$\!$ dependence, which was
mentioned above many times. But here it is important to note that the assumption about the power-law behavior
of $D\left(k\right)$ for large $k$ is unnecessary. In other words, in the framework of the approaches based on
the FRG and HRG methods, there is no assumption about the power-law tail contained in the perturbative
RG-theory of turbulence, and precisely this assumption is a weak point in the perturbative RG-theory.

Now we denote the metric tensor element by $\bar{g}_{\boldsymbol{r}}=g$. In what follows, it is important that
$g\left(0\right)=\infty$ and $g\left(\infty\right)=0$. These requirements are satisfied, for example, for the
metric tensor of the AdS space. We need the following change for the pumping function:
\begin{eqnarray}
D\left(k\right)\rightarrow D_{\varLambda\left(z\right)}\left(k\right)=
D\left(\frac{k}{\sqrt{g\left(z\right)}}\right).\label{sep6}
\end{eqnarray}
The asymptotics of expression (\ref{sep6}) as $g\rightarrow\infty$ has the form
\begin{eqnarray}
D_{\varLambda\left(z\right)}\left(k\right)\big|_{g\rightarrow\infty}=
D_{0}\,g^{-\frac{\eta_{d}}{2}}\left(\frac{k}{m}\right)^{\eta_{d}}.\label{sep7}
\end{eqnarray}

Now we introduce the inverse pumping function $N$ and also the actual variable~$\chi$:
\begin{eqnarray}
N_{\varLambda\left(z\right)}\left(k\right)=\frac{1}{D_{\varLambda\left(z\right)}
\left(k\right)}=N_{0}\left(\frac{k}{\sqrt{g\left(z\right)}m}\right)^{-\eta_{d}}
\chi_{\varLambda\left(z\right)}\left(k\right).\label{sep8}
\end{eqnarray}
Thus, $N$ is used to choose the function $\chi$ as follows:
\begin{eqnarray}
\chi_{\varLambda\left(z\right)}\left(k\right)=\frac{1}{d\left(\frac{k}
{\sqrt{g\left(z\right)}m}\right)}.\label{sep9}
\end{eqnarray}
Expressions (\ref{sep8})--(\ref{sep9}) imply the following important conceptual aspect of the entire paper from
the holographic standpoint: up to certain "details"$\!$, the pumping function becomes an actual variable, i.e.,
it becomes an argument of the solution! Such a "combination"$\!$ allows us to obtain interesting answers for
different Green's functions. Here we note that the pumping function, which plays the role of the amplitude of
the solution (such an HRG-flow can be constructed), does not generate such answers.

In what follows, we need a model more general than expression (\ref{sep8}). The desired generalization states
\begin{eqnarray}
N_{\varLambda\left(z\right)}\left(k\right)=N_{0,\varLambda\left(z\right)}
g^{\frac{\eta_{d}}{2}}\left(z\right)\left(\frac{k}{m}\right)^{-\eta_{d}}
\chi_{\varLambda\left(z\right)}\left(k\right).\label{sep10}
\end{eqnarray}
As previously, for the function $\chi$ to be an actual variable, it suffices to satisfy the condition:
\begin{eqnarray}
N_{0,\varLambda\left(z\right)}g^{\frac{\eta_{d}}{2}}\left(z\right)=N_{0}.\label{sep11}
\end{eqnarray}
In this case, we immediately determine several values for our HRG-flow:
\begin{eqnarray}
\chi_{\varLambda\left(z=0\right)}\left(k\right)=\chi\left(k\right)=1,\quad\eta_{n}=1.\label{sep12}
\end{eqnarray}
It remains to make some assumptions about the viscosity $\bar{\nu}$. We assume that $\bar{\nu}$ depends on $k$
only in terms of the actual variable $\chi$:
\begin{eqnarray}
\bar{\nu}_{\chi}\left(k\right)=\bar{\nu}_{\chi}=\nu\chi^{\eta_{\nu}}.\label{sep13}
\end{eqnarray}
Such a choice is natural after the minimal choice, where $\bar{\nu}$ is a constant, and also agrees with
expression (\ref{extmod1}). The exponent $\eta_{\nu}$ is a free parameter of the problem. Finally, for
convenience and for the compactness of the further expressions, we make the change $\eta_{\nu}\rightarrow
-\eta_{\nu}$.

Under all the above assumptions, we can now write the solution (\ref{sol7})--(\ref{sol8}) for the two-particle
PGF as follows:
\begin{eqnarray}
\mathcal{G}\left(\omega,k\right)=Ck^{4-\eta_{d}}
\left\{\sigma+\frac{2\eta_{\nu}x\left(\omega,k\right)I_{\frac{\xi}{2\eta_{\nu}}+1}
\left[x\left(\omega,k\right)\right]}{I_{\frac{\xi}{2\eta_{\nu}}}
\left[x\left(\omega,k\right)\right]}\right\}.\label{sep14}
\end{eqnarray}
And here we also write all the "construction blocks"$\!$ of PGF (\ref{sep14}):
\begin{eqnarray}
x\left(\omega,k\right)=-\frac{\sqrt{\rho}}{\eta_{\nu}}\frac{|\omega|}
{\nu k^{2}},\quad C=\frac{\nu^{2}N_{0}m^{\eta_{d}}}{2\rho},\nonumber\\
\sigma=\eta-1+2\eta_{\nu}+\xi,\quad\xi=
\sqrt{\left(1-2\eta_{\nu}-\eta\right)^{2}+4\rho}.\label{sep15}
\end{eqnarray}
Expressions (\ref{sep14})--(\ref{sep15}) explicitly present the final solution of our problem. They describe
the holographic version of PGF satisfying all the requirements stated above.

Let us consider an example of PGF (\ref{sep14})--(\ref{sep15}) for given values of the parameters of the model.
Let $\eta=0$, and let $\xi=2\eta_{\nu}$ (the last assumption corresponds to the modified first- and
second-order Bessel functions). In this case, $\rho=\eta_{\nu}-1/4$ and $\sigma=4\rho$. Now we set
$\eta_{\nu}=1/3$. Expressions (\ref{sep14})--(\ref{sep15}) become
\begin{eqnarray}
\mathcal{G}\left(\omega,k\right)=\frac{C}{3}\,k^{3}
\left\{1+2\,\frac{x\left(\omega,k\right)I_{2}\left[x\left(\omega,k\right)\right]}
{I_{1}\left[x\left(\omega,k\right)\right]}\right\},\quad x\left(\omega,k\right)=
-\frac{\sqrt{3}}{2}\frac{|\omega|}{\nu k^{2}}.\label{sep16}
\end{eqnarray}
Here we also used the additional equality $\eta_{d}=4-D=1$ in the case of three-dimensional
$\boldsymbol{r}$-space.

It is time to discuss the meaning of the boundary action $\mathcal{G}$ and the BGF family generated by this
action. As the starting point, we discuss expression (\ref{sol7}) for the two-particle PGF. The latter implies
that the two-particle PGF has the meaning of the self-energy $\varSigma\left(\omega,k\right)$, and hence of a
one-particle irreducible vertex function with two external "legs"$\!$. The corresponding two-particle BGF,
expressed by formulas (\ref{sol1})--(\ref{sol4}), has the same meaning. This coincidence is not accidental.

As an example, the AdS/CFT correspondence shows that if the latter is true, then the boundary action
$\mathcal{G}$ is a functional Legendre transformation of the Wilsonian effective action \cite{LMPV}. The latter
satisfies the Wilson--Polchinski equation, which is also satisfied by the ACGF generating functional expressed,
after a simple change of variables, in terms of the generating functional of connected Green's functions
$\mathcal{G}_{c}$ \cite{KBS}. We can make an intermediate conclusion that, on the "qualitative"$\!$ level, the
boundary action $\mathcal{G}$ is a functional Legendre transformation of the generating functional of connected
Green's functions $\mathcal{G}_{c}$. On the other hand, we know that the functional Legendre transformation of
the quantity $\mathcal{G}_{c}$ is the generating functional of one-particle irreducible vertices
$\varGamma_{1\mathrm{PI}}$. Thus, on the "qualitative"$\!$ level, the boundary action $\mathcal{G}$ coincides
with the  generating functional of $\varGamma_{1\mathrm{PI}}$. And the qualitative corre\-spon\-dence between
the functionals, i.e., families of Green's functions (irreducible vertices), can only be realized in a certain
limit. In the general case, there is a (nonperturbative) difference, for example, between the two-particle PGF
and the exact self-energy which is equal to $\varDelta\varSigma\left(\omega,k\right)$.

Further, we assume that the FRG-theory of turbulence is formulated in terms of one-particle irreducible vertex
functions following the Wetterich--Morris formalism \cite{Wetterich}--\cite{Morris}. The Wetterich--Morris
approach is based on the abstract FRG-flow equation (for most applications, the momentum-frequency FRG-flow
equation) for the $\varLambda$-dependent generating functional of $\varGamma_{1\mathrm{PI}}$ (strictly
speaking, with a certain distinction from such a functional, because the second functional derivative in the
Wetterich--Morris approach generates the inverse propagator rather than the self-energy). As usual, the
functional equation is used to derive the hierarchy of integro-differential equations for irreducible vertices,
which contains the $n,n+1$ problem. With the above discussion of the "qualitative"$\!$ coincidence between
$\mathcal{G}$ and $\varGamma_{1\mathrm{PI}}$ taken into account, we conclude that such a hierarchy must
approximately be satisfied for the BGF family. But we can again play a different game, i.e., we can verify
whether the FRG hierarchy is satisfied, for example, if we assume that the solution is exact for the
two-particle BGF.

In the light of the above discussion, we consider another important problem, i.e., the relationship between the
FRG-regulator $R_{\varLambda}$ and the additional coordinates of the curved space, for example, the coordinate
$z$. And here relations (\ref{sol1})--(\ref{sol4}) for the two-particle BGF suggest how to act. For example,
the AdS/CFT correspondence implies that the two theories must coincide at the level of correlation functions.
This coincidence can be treated either as approximate, like in the original AdS/CFT version, or as exact for
some functions, for example, for two-particle functions. For definiteness, let us consider the second version.
We know that the values $z=0$ and $\varLambda=0$ generate the "physical"$\!$ theory. In the opposite limit case
$z\rightarrow\infty$ and $\varLambda\rightarrow\infty$, the expressions for the two-point BGF and the
self-energy become infinite, just as this must be. Taking all this into account, we can make an important
conclusion that, in the simplest case, the quantities $R_{\varLambda}$ and $z$ are proportional (the
proportionality coefficient is a dimensional quantity). Of course, this is the simplest scenario. In a more
general case, such a relationship can be treated as an additional "degree of freedom"$\!$ and be chosen so as
to satisfy some requirements imposed on the functional flow.

This completes our analysis of the solution of the HRG-flow equation (\ref{extmod16}) for the two-particle BGF
corresponding to the "standard quantization"$\!$ case, in particular, our analysis of the corresponding
two-particle PGF. In the subsequent subsection, we briefly illustrate the material considered above with an
example of the second independent solution of equation (\ref{extmod16}) corresponding to the "alternative
quantization"$\!$ and then consider some additional problems of the HRG-theory of turbulence.

\subsection{Alternative quantization}

We briefly consider the solution of equation (\ref{extmod16}) for the two-particle BGF, which is an analog of
the "alternative quantization"$\!$ in the AdS space (in the general solution, the constant of integration is
$C_{\mathcal{G}}=\infty$). This solution is similar to expression (\ref{sol1}):
\begin{eqnarray}
\mathcal{G}_{\chi}\left(\omega,k\right)=\frac{N_{\chi}\left(k\right)\bar{\nu}_{\chi}^{2}
\left(k\right)k^{4}}{2\rho}\left\{\eta-\eta_{n}-2\eta_{\nu}+2\chi\partial_{\chi}
\ln K_{\frac{\xi}{2\eta_{\nu}}}\left[x_{\chi}\left(\omega,k\right)\right]\right\}.\label{besselk1}
\end{eqnarray}
The solution (\ref{besselk1}) can also be represented in the form which does not contain any derivative if we
use the relation
\begin{eqnarray}
\chi\partial_{\chi}\ln K_{\frac{\xi}{2\eta_{\nu}}}\left[x_{\chi}
\left(\omega,k\right)\right]=
-\frac{\xi}{2}+\frac{\eta_{\nu}x_{\chi}\left(\omega,k\right)
K_{\frac{\xi}{2\eta_{\nu}}+1}\left[x_{\chi}\left(\omega,k\right)\right]}
{K_{\frac{\xi}{2\eta_{\nu}}}\left[x_{\chi}\left(\omega,k\right)\right]}.\label{besselk2}
\end{eqnarray}
In the last two expressions, we introduced the notation
\begin{eqnarray}
x_{\chi}\left(\omega,k\right)=\frac{\sqrt{\rho}}{\eta_{\nu}}\frac{|\omega|}
{\bar{\nu}_{\chi}\left(k\right)k^{2}},\quad
\xi=\sqrt{\left(\eta-\eta_{n}-2\eta_{\nu}\right)^{2}+4\rho}.\label{besselk3}
\end{eqnarray}
Formulas (\ref{besselk1})--(\ref{besselk3}) completely describe the second solution of equation
(\ref{extmod16}).

Now it is easy to formulate the corresponding two-particle PGF defined on the boundary of the curved space
$z=0$. The expression for such a function reads
\begin{eqnarray}
\mathcal{G}\left(\omega,k\right)=\frac{N_{B}\left(k\right)\bar{\nu}_{B}^{2}\left(k\right)k^{4}}{2\rho}
\left\{\sigma'+\frac{2\eta_{\nu}x\left(\omega,k\right)K_{\frac{\xi}{2\eta_{\nu}}+1}
\left[x\left(\omega,k\right)\right]}{K_{\frac{\xi}{2\eta_{\nu}}}
\left[x\left(\omega,k\right)\right]}\right\}.\label{besselk4}
\end{eqnarray}
Now we also introduce the notation
\begin{eqnarray}
x\left(\omega,k\right)=\frac{\sqrt{\rho}}{\eta_{\nu}}\frac{|\omega|}{\bar{\nu}_{B}
\left(k\right)k^{2}},\quad\sigma'=\eta-\eta_{n}-2\eta_{\nu}-\xi.\label{besselk5}
\end{eqnarray}
Further we use the assumptions made to choose the random force correlator (in this case, the sign of
$\eta_{\nu}$ remains the same). The solution (\ref{besselk4})--(\ref{besselk5}) becomes
\begin{eqnarray}
\mathcal{G}\left(\omega,k\right)=Ck^{4-\eta_{d}}
\left\{\sigma'+\frac{2\eta_{\nu}x\left(\omega,k\right)K_{\frac{\xi}{2\eta_{\nu}}+1}
\left[x\left(\omega,k\right)\right]}{K_{\frac{\xi}{2\eta_{\nu}}}
\left[x\left(\omega,k\right)\right]}\right\}.\label{besselk6}
\end{eqnarray}
The notation in expression (\ref{besselk6}) is
\begin{eqnarray}
x\left(\omega,k\right)=\frac{\sqrt{\rho}}{\eta_{\nu}}\frac{|\omega|}
{\nu k^{2}},\quad C=\frac{\nu^{2}N_{0}m^{\eta_{d}}}{2\rho},\nonumber\\
\sigma'=\eta-1-2\eta_{\nu}-\xi,\quad
\xi=\sqrt{\left(1+2\eta_{\nu}-\eta\right)^{2}+4\rho}.\label{besselk7}
\end{eqnarray}
Formulas (\ref{besselk6})--(\ref{besselk7}) describe the holographic version of PGF corresponding to the second
solution of the HRG-flow equation (\ref{extmod16}). This quantity must also have the meaning of some observable
correlation function of the velocity and hence be compared with experimental data and with the predictions made
in other theories, in particular, in the perturbative RG-theory of turbulence. This completes our analysis of
the solutions of the HRG-flow equation (\ref{extmod16}). In the following subsection, we briefly consider a
certain generalization of the HRG-flow constructed in this paper.

\subsection{Total derivative with respect to the scale}

The simple (at the first glance) derivative with respect to the flow variable~$z$ in the left-hand side of the
HJ functional equation (\ref{2hrgtt15}) is an amazingly interesting element of construction of various
holographic models. Such a derivative is an example of the derivative with respect to the scale of the theory.
To preserve the generality of our consideration, we here use the notion of scale, which is denoted
by~$\varLambda$. Moreover, such a general approach permits regarding the idea of not only holographic but also
exact FRG-flow from "one more"$\!$ standpoint. In the framework of this approach, we deal with the functional
flow equations for $\varLambda$-dependent generating functionals of some family of Green's functions.

Now we consider the "generalized"$\!$ flow, which means that the derivative with respect to the scale
$\partial_{\varLambda}$ is replaced by a more general structure, i.e., by a certain "total"$\!$ derivative of
the form
\begin{eqnarray}
\partial_{\varLambda}\rightarrow \hat{\partial}_{\varLambda}=\partial_{\varLambda}+
\mathcal{F}^{\left(I\right)}_{\varLambda,a}\left[\boldsymbol{v},V,\ldots\right]
\frac{\delta}{\delta v_{a}}+\mathcal{F}^{\left(II\right)}_{\varLambda,ab}
\left[\boldsymbol{v},V,\ldots\right]\frac{\delta}{\delta V_{ab}}+\ldots\label{totder1}
\end{eqnarray}
Expression (\ref{totder1}) is written in terms of superindices, the field configurations $v_{a}$ and $V_{ab}$
with one and two indices (the further generalization is obvious) are functional variables, and the quantities
$\mathcal{F}^{\left(I\right)}$ and $\mathcal{F}^{\left(II\right)}$ are some $\varLambda$-dependent functionals.
The functional generalization of the "kinetic coefficient"$\!$ can easily be taken into account in
(\ref{totder1}):
\begin{eqnarray}
\hat{\partial}_{\varLambda}\rightarrow\varGamma_{\varLambda}
\left[\boldsymbol{v},V,\ldots\right]\hat{\partial}_{\varLambda}.\label{totder2}
\end{eqnarray}
We consider the simplest realization of the model (\ref{totder1})--(\ref{totder2}). This version contains only
the velocity field $\boldsymbol{v}$:
\begin{eqnarray}
\mathcal{F}^{\left(I\right)}_{\varLambda,a}\left[\boldsymbol{v},V,\ldots\right]=
F_{\varLambda,ab}v_{b},\quad\mathcal{F}^{\left(II\right)}_{\varLambda,ab}
\left[\boldsymbol{v},V,\ldots\right]=\mathcal{F}^{\left(III\right)}_{\varLambda,ab}
\left[\boldsymbol{v},V,\ldots\right]=\ldots=0.\label{totder3}
\end{eqnarray}
\begin{eqnarray}
\varGamma_{\varLambda}\left[\boldsymbol{v},V,\ldots\right]=\varGamma_{\varLambda}.\label{totder4}
\end{eqnarray}
The model (\ref{totder3})--(\ref{totder4}) has the following important property: it does not contain the
$n,n+1$ problem. Thus, if we apply such a generalization to the HJ equation, then the chain of
integro-differential equations for Green's functions due to this generalization remains "uncoupled"$\!$. An
example of the flow (\ref{totder3})--(\ref{totder4}) for a certain choice of $F$ is the case of
$\varLambda$-dependent field configurations $\boldsymbol{v}$ \cite{HGCW} (in the FRG).

Now we pass from abstract superindices to a specific representation. As the first step, we choose the
$\left(z,x\right)$-representation. Moreover, in this paper, we restrict our consideration to (operators) $F$
such that the terms in expression (\ref{totder1}), which are complementary to the derivative
$\partial_{\varLambda}$, become
\begin{eqnarray}
\varDelta_{F}^{\left(0\right)}\left(z\right)=\int d^{D+1}x\,v_{\alpha}\left(z,x\right)
F_{\varLambda\left(z\right),\alpha\beta}\left(\partial_{x}\right)\frac{\delta}{\delta v_{\beta}
\left(z,x\right)}.\label{totder5}
\end{eqnarray}
\begin{eqnarray}
\varDelta_{F}^{\left(1\right)}\left(z\right)=\sum\limits_{i=1}^{D+1}\int d^{D+1}x
\left[x^{i}\partial_{x^{i}}v_{\alpha}\left(z,x\right)\right]F^{\left(i\right)}_{\varLambda
\left(z\right),\alpha\beta}\left(\partial_{x}\right)\frac{\delta}{\delta v_{\beta}
\left(z,x\right)}.\label{totder6}
\end{eqnarray}
In connection with expression (\ref{totder6}), we note that the operators $F^{\left(i\right)}$ coincide for
$i=1\ldots D$. We use the form containing (\ref{totder6}) exclusively for convenience.

In the case of operators (\ref{totder5})--(\ref{totder6}), the left-hand side (\ref{2hrgtt21}) of the equation
for the two-particle BGF acquires the following additional terms after renormalization with respect to
$\delta\left(0\right)$:
\begin{eqnarray}
\varDelta^{\left(0\right)}\mathrm{lhs}=F_{\varLambda\left(z\right),\alpha_{1}\gamma}
\left(\partial_{x_{1}}\right)\mathcal{G}^{\left(2\right)}_{\varLambda\left(z\right),\gamma\alpha_{2}}
\left(z;x_{1},x_{2}\right)+\left(1\rightleftarrows2\right).\label{totder7}
\end{eqnarray}
\begin{eqnarray}
\varDelta^{\left(1\right)}\mathrm{lhs}=-\sum\limits_{i=1}^{D+1}\partial_{x^{i}_{1}}
\left[x^{i}_{1}F^{\left(i\right)}_{\varLambda\left(z\right),\alpha_{1}\gamma}
\left(\partial_{x_{1}}\right)\mathcal{G}^{\left(2\right)}_{\varLambda\left(z\right),\gamma\alpha_{2}}
\left(z;x_{1},x_{2}\right)\right]+\left(1\rightleftarrows2\right).\label{totder8}
\end{eqnarray}

If the system is translation invariant with respect to the coordinates $\left(t,\boldsymbol{r}\right)$, then
the terms (\ref{totder7})--(\ref{totder8}) in the momentum-frequency representation take the simplest form
\begin{eqnarray}
\varDelta^{\left(0\right)}\mathrm{lhs}=2F_{\varLambda\left(z\right),\alpha_{1}\gamma}
\left(\boldsymbol{k}\right)\mathcal{G}_{\varLambda\left(z\right),\gamma\alpha_{2}}
\left(z;\omega,\boldsymbol{k}\right).\label{totder9}
\end{eqnarray}
\begin{eqnarray}
\varDelta^{\left(1\right)}\mathrm{lhs}=\sum\limits_{i=1}^{D+1}\left[-1+k^{i}\partial_{k^{i}}\right]
F^{\left(i\right)}_{\varLambda\left(z\right),\alpha_{1}\gamma}
\left(\boldsymbol{k}\right)\mathcal{G}_{\varLambda\left(z\right),\gamma\alpha_{2}}
\left(z;\omega,\boldsymbol{k}\right).\label{totder10}
\end{eqnarray}
Projecting expressions (\ref{totder9})--(\ref{totder10}) on the transverse direction of the momentum
$\boldsymbol{k}$, we obtain the corresponding additional contributions to the left-hand side (\ref{2hrgtt31})
of the transverse flow equation for the two-particle BGF.

Now we discuss the properties of the contributions (\ref{totder9})--(\ref{totder10}). An example of such
contributions for a certain choice of the values of $F$ is the case of rescaled flow equations. The rescaled
equations have been considered many times in the literature. Further, the term (\ref{totder9}) can be used to
obtain the elongated derivative, which was already encountered in connection with equation (\ref{extmod16}).
But in the general case, the contribution (\ref{totder10}) can transform the ordinary differential equation
into a partial differential equation and thus significantly change the nature of the final equation which,
consequently, changes the character of the solutions. In the general case, the presence of contributions
(\ref{totder9})--(\ref{totder10}) is a part of the statement of the problem, and each substitution has its own
justification.

In the framework of the discussion of the ambiguity of the RG-flow procedures, we pay our attention to another
scenario but now from the FRG side. The Wilson--Polchinski abstract FRG-flow equation (\ref{3frgstt16}) is a
special case of a more general functional flow structure (we preserve the notation introduced in the first part
of the paper) \cite{Ros}:
\begin{eqnarray}
\partial_{\varLambda}\mathcal{G}_{\varLambda}\left[\boldsymbol{u}\right]=
\left(\mathcal{G}^{\left(1\right)}_{\varLambda}\left[\boldsymbol{u}\right]\Big|
\varPsi_{\mathcal{G},\varLambda}\left[\boldsymbol{u}\right]\right)
+\mathrm{Tr}\left(\varPsi^{\left(1\right)}_{\mathcal{G},\varLambda}
\left[\boldsymbol{u}\right]\right).\label{rosten1}
\end{eqnarray}
Abstract equation (\ref{rosten1}) can be rewritten in terms of superindices as
\begin{eqnarray}
\partial_{\varLambda}\mathcal{G}_{\varLambda}\left[\boldsymbol{u}\right]=
\mathcal{G}^{\left(1\right)}_{\varLambda,a}\left[\boldsymbol{u}\right]
\varPsi_{\mathcal{G},\varLambda,a}\left[\boldsymbol{u}\right]+
\varPsi^{\left(1\right)}_{\mathcal{G},\varLambda,aa}\left[\boldsymbol{u}\right].\label{rosten2}
\end{eqnarray}
The object $\varPsi$ has one superindex $a$ and is a functional of the field variable $\boldsymbol{u}$.
Moreover, it depends on the functional $\mathcal{G}$, which makes the functional flow equation
(\ref{rosten1})--(\ref{rosten2}) nonlinear. The meaning of $\varPsi$ is that this object parameterizes some
process which increases the level of the coarse-grained nature of the degree of freedom, in other words, some
FRG-flow procedure. In this case, $\varPsi$ satisfies only the most general requirements, and its specific form
is an element of the statement of the problem.

To obtain the Wilson--Polchinski equation from (\ref{rosten1})--(\ref{rosten2}), we must perform the following
\textbf{choice} (we note that there is (first term of) an expansion in a power series in the functional
$\mathcal{G}$) \cite{Ros}:
\begin{eqnarray}
\varPsi_{\mathcal{G},\varLambda}\left[\boldsymbol{u}\right]=\frac{1}{2}
\left[\partial_{\varLambda}\mathbf{G}_{0,\varLambda}\right]
\mathcal{G}^{\left(1\right)}_{\varLambda}\left[\boldsymbol{u}\right].\label{rosten3}
\end{eqnarray}
In terms of superindices, expression (\ref{rosten3}) becomes
\begin{eqnarray}
\varPsi_{\mathcal{G},\varLambda,a}\left[\boldsymbol{u}\right]=
\frac{1}{2}\left[\partial_{\varLambda}\mathbf{G}_{0,\varLambda}\right]_{ab}
\mathcal{G}^{\left(1\right)}_{\varLambda,b}\left[\boldsymbol{u}\right].\label{rosten4}
\end{eqnarray}
Formulas (\ref{rosten1})--(\ref{rosten4}) demonstrate a wide functional ambiguity of the FRG method. But this
ambiguity is not a drawback for the following reason. Our final goal is to obtain the complete phase diagram of
the theory under study (i.e., the set of all fixed points and possibly "something else"$\!$). If a fixed point
is inaccessible in the framework of one FRG-flow procedure, then it can in principle be accessible in the
framework of another FRG-flow procedure. Of course, everything depends on the concrete theory, but the above
consideration allows us to hope that, in the framework of the FRG method, it is possible to present the
"complete set"$\!$ of flow procedures, which permits constructing the complete phase diagram of the theory.

\section{Synthesis of the FRG and HRG methods\\ in an example of the simplest solutions}\label{p4}

This material will be available in the second version of this paper.

\section{Conclusion}\label{p5}

In this paper, we consider the system of hydrodynamic equations for incompressible viscous fluid which consists
of the stochastic Navier--Stokes equation, the continuity equation, and the equation of state (in the case of
incompressible fluid, the latter is trivial and means that the density is constant). For this system, we
construct a nonlocal quantum-field model. This model is then investigated by the methods of functional and
holographic renormalization groups.

In the framework of the FRG method, we define the $\varLambda$-deformed analogs of the quantities participating
in the nonlocal quantum-field model. In particular, we introduce $\varLambda$-dependent ACGFs and the
Wilson--Polchinski abstract FRG-flow equation, which is satisfied by the corresponding generating functional.
The FRG-flow equation for the two-particle Green's function is derived in the momentum-frequency
representation.

The obtained equation contains the projected four-particle ACGF in a special kinematics, and this equation is
therefore nonclosed. This $n,n+1$ problem reflects the entire difficulty of the FRG method. For this reason, an
important role is played by additional relations which permit obtaining more information about the ACGF and
other families of Green's functions. An example of such relations is the hierarchy of integro-differential
equations generated by the Ward functional identity for the time-gauged Galilean symmetry. A remarkable
property of this hierarchy is the fact that it does not contain the $n,n+1$ problem. In the momentum-frequency
representation, such a hierarchy permits expressing higher-order functions in a special kinematics in terms of
lower-order functions.

An example of such relations, which is more complicated from the conceptual standpoint, is the hierarchy of
equations for the BGFs generated by the boundary action, which satisfies the HJ functional equation in the HRG
method. Starting from the HJ equation, the HRG-flow equation for the two-particle boundary Green's function is
derived in the momentum-frequency representation. Because the HJ equation does not contain the $n,n+1$ problem,
the equation for the two-particle BGF is closed and is a generalization of a special Riccati equation.

In the framework of the minimal holographic model and its simple generalization, we obtained an explicit
analytic solution of such an equation, which is an interesting combination of the modified Bessel functions $I$
and $K$. An important property of the obtained solution is the fact that it minimally depends on the details of
the function of the energy pumping into the system, which models the stochasticity.

The restrictions imposed on the RG-flow models which are related to the time-gauged Galilean symmetry, as well
as the problem of choosing the pumping function and some generalizations of the RG-flow procedures applied in
the paper, are considered in detail both from the HRG and FRG sides. Finally, we study the possibility of using
the HRG-solutions to construct solutions of the FRG-flow equation for the two-particle Green's function.

In conclusion, we point out another positive detail. As previously noted, the hierarchy of FRG-flow equations
is the most complicated mathematical object because of the $n,n+1$ coupling. The HRG-hierarchy can therefore be
considered as a separate and independent simplified structure which allows us not only to improve our
understanding of the general properties of the functional flow but also to obtain explicit analytic solutions
for different families of Green's functions. Of course, to analyze the results obtained by the HRG method
through the FRG "prism"$\!$ is one of the most important goals in subsequent papers.

\section*{Acknowledgments}
The author thanks to N.~V.~Antonov for the useful discussions and the advise.

%
%
%


\end{document}